\documentclass[twocolumn, trackchanges]{aastex62}

\submitjournal{ApJ}
\usepackage{amsmath}
\usepackage{comment}
\usepackage{cleveref}

\definecolor{SecureBlue}{HTML}{E3F2FD}   
\definecolor{ProbYellow}{HTML}{FFF9C4}   
 \usepackage[table]{xcolor}

\crefformat{section}{\S#2#1#3} 
\crefformat{subsection}{\S#2#1#3}
\crefformat{subsubsection}{\S#2#1#3}

\shortauthors{Rinaldi et al.}

\begin{document}

\title{\bf The Way We Tally Becomes the Tale: the Impact of Selection Strategies on the Inferred Evolution of Little Red Dots Across Cosmic Time}

\newcommand{\gsim}{{\;\raise0.3ex\hbox{$>$\kern-0.75em\raise-1.1ex\hbox{$\sim$}}\;}}

\correspondingauthor{Pierluigi Rinaldi}
\email{prinaldi@stsci.edu}

\author[0000-0002-5104-8245]{Pierluigi Rinaldi}
\affiliation{Space Telescope Science Institute, 3700 San Martin Drive, Baltimore, Maryland 21218, USA}
\author[0000-0003-4565-8239]{Kevin Hainline}
\affiliation{Steward Observatory, University of Arizona, 933 North Cherry Avenue, Tucson, AZ 85721, USA}

\author[0000-0003-2388-8172]{Francesco D'Eugenio}
\affiliation{Kavli Institute for Cosmology, University of Cambridge, Madingley Road, Cambridge, CB3 0HA, UK}
\affiliation{Cavendish Laboratory, University of Cambridge, 19 JJ Thomson Avenue, Cambridge, CB3 0HE, UK}

\author[0000-0003-4528-5639]{Pablo G. P\'erez-Gonz\'alez}
\affiliation{Centro de Astrobiolog\'ia (CAB), CSIC–INTA, Cra. de Ajalvir Km.~4, 28850- Torrej\'on de Ardoz, Madrid, Spain}

\author[0000-0002-2929-3121]{Daniel J.\ Eisenstein}
\affiliation{Center for Astrophysics $|$ Harvard \& Smithsonian, 60 Garden St., Cambridge MA 02138 USA}

\author[0000-0001-9262-9997]{Christopher N. A. Willmer}
\affiliation{Steward Observatory, University of Arizona, 933 North Cherry Avenue, Tucson, AZ 85721, USA}

\author[0000-0001-6301-3667]{Courtney Carreira}
\affiliation{Department of Astronomy and Astrophysics, University of California, Santa Cruz, 1156 High Street, Santa Cruz, CA 95064, USA}

\author[0000-0002-4271-0364]{Brant Robertson}
\affiliation{Department of Astronomy and Astrophysics, University of California, Santa Cruz, 1156 High Street, Santa Cruz, CA 95064, USA}

\author[0000-0002-9280-7594]{Benjamin D.\ Johnson}
\affiliation{Center for Astrophysics $|$ Harvard \& Smithsonian, 60 Garden St., Cambridge MA 02138 USA}

\author[0000-0002-8909-8782]{Stacey Alberts}
\affiliation{AURA for the European Space Agency (ESA), Space Telescope Science Institute, 3700 San Martin Dr., Baltimore, MD 21218, USA}

\author[0000-0003-0215-1104]{William M. Baker}
\affiliation{DARK, Niels Bohr Institute, University of Copenhagen, Jagtvej 155A, DK-2200 Copenhagen, Denmark}

\author[0000-0002-8651-9879]{Andrew J. Bunker}
\affiliation{Department of Physics, University of Oxford, Denys Wilkinson Building, Keble Road, Oxford OX13RH, UK}

\author[0000-0002-6719-380X]{Stefano Carniani}
\affiliation{Scuola Normale Superiore, Piazza dei Cavalieri 7, I-56126 Pisa, Italy}

\author[0000-0003-1344-9475]{Eiichi Egami}
\affiliation{Steward Observatory, University of Arizona, 933 North Cherry Avenue, Tucson, AZ 85721, USA}

\author[0000-0003-4337-6211]{Jakob M.\ Helton}
\affiliation{Department of Astronomy \& Astrophysics, The Pennsylvania State University, University Park, PA 16802, USA}

\author[0000-0001-7673-2257]{Zhiyuan Ji}
\affiliation{Steward Observatory, University of Arizona, 933 North Cherry Avenue, Tucson, AZ 85721, USA}

\author[0009-0003-7423-8660]{Ignas Juod\v{z}balis}
\affiliation{Kavli Institute for Cosmology, University of Cambridge, Madingley Road, Cambridge, CB3 0HA, UK}
\affiliation{Cavendish Laboratory, University of Cambridge, 19 JJ Thomson Avenue, Cambridge, CB3 0HE, UK}

\author[0000-0001-6052-4234]{Xiaojing Lin}
\affiliation{Department of Astronomy, Tsinghua University, Beijing 100084, China}
\affiliation{Steward Observatory, University of Arizona, 933 North Cherry Avenue, Tucson, AZ 85721, USA}

\author[0000-0002-6221-1829]{Jianwei Lyu}
\affiliation{Steward Observatory, University of Arizona, 933 North Cherry Avenue, Tucson, AZ 85721, USA}

\author[0009-0003-5402-4809]{Zheng Ma}
\affiliation{Steward Observatory, University of Arizona, 933 North Cherry Avenue, Tucson, AZ 85721, USA}

\author[0000-0002-4985-3819]{Roberto Maiolino}
\affiliation{Kavli Institute for Cosmology, University of Cambridge, Madingley Road, Cambridge, CB3 0HA, UK}
\affiliation{Cavendish Laboratory, University of Cambridge, 19 JJ Thomson Avenue, Cambridge, CB3 0HE, UK}
\affiliation{Department of Physics and Astronomy, University College London, Gower Street, London WC1E 6BT, UK}

\author[0000-0002-7392-7814]{Eleonora Parlanti}

\affiliation{Scuola Normale Superiore, Piazza dei Cavalieri 7, I-56126 Pisa, Italy}

\author[0000-0001-6010-6809]{Jan Scholtz}
\affiliation{Kavli Institute for Cosmology, University of Cambridge, Madingley Road, Cambridge, CB3 0HA, UK}
\affiliation{Cavendish Laboratory, University of Cambridge, 19 JJ Thomson Avenue, Cambridge, CB3 0HE, UK}

\author[0000-0001-6561-9443]{Yang Sun}
\affiliation{Steward Observatory, University of Arizona, 933 North Cherry Avenue, Tucson, AZ 85721, USA}

\author[0000-0002-8224-4505]{Sandro Tacchella}
\affiliation{Kavli Institute for Cosmology, University of Cambridge, Madingley Road, Cambridge, CB3 0HA, UK}
\affiliation{Cavendish Laboratory, University of Cambridge, 19 JJ Thomson Avenue, Cambridge, CB3 0HE, UK}

\author[0000-0001-8349-3055]{Giacomo Venturi}
\affiliation{Scuola Normale Superiore, Piazza dei Cavalieri 7, I-56126 Pisa, Italy}

\author[0000-0003-2919-7495]{Christina C. Williams}
\affiliation{NSF–DOE Vera C. Rubin Observatory/NSF NOIRLab, 950 N. Cherry Ave., Tucson, AZ 85719, USA}

\author[0000-0002-4201-7367]{Chris Willott}
\affiliation{NRC Herzberg, 5071 West Saanich Rd, Victoria, BC V9E 2E7, Canada}

\author[0000-0002-7595-121X]{Joris Witstok}
\affiliation{Cosmic Dawn Center (DAWN), Copenhagen, Denmark}
\affiliation{Niels Bohr Institute, University of Copenhagen, Jagtvej 128, DK-2200, Copenhagen, Denmark}

\author[0000-0002-8876-5248]{Zihao Wu}
\affiliation{Center for Astrophysics $|$ Harvard \& Smithsonian, 60 Garden St., Cambridge MA 02138 USA}

\begin{abstract}

Little Red Dots (LRDs) have emerged as a key population linked to early black hole growth, yet photometric selections have predominantly targeted only the most extreme red systems, thereby shaping our current understanding of this new population of objects. In this work, we deliberately explore a broad range of optical redness while enforcing stringent compactness and visual inspection to ensure robustness and minimize contamination. Leveraging the depth and multiwavelength coverage of the JWST Advanced Deep Extragalactic Survey (JADES) data in the GOODS-North and GOODS-South fields, we construct the largest photometric census of LRDs to date in these fields, comprising 412 sources over $z\approx2\text{--}11$ across $\approx349.6$ arcmin$^2$. We show that classic extreme color cuts isolate only a minor fraction of this population ($\lesssim25\%$), while the majority of LRDs span a broader, largely unexplored parameter space. We quantify how selection strategies impact UV and optical luminosity functions and number density evolution, finding that current demographic trends of LRDs are strongly driven by selection biases and further limited by incomplete identification at both high and low redshift. Spectroscopically confirmed LRDs reveal a continuous range of spectral shapes, consistent with varying Active Galactic Nucleus (AGN) and host contributions in agreement with recent findings. Our results demonstrate that commonly adopted, purity-driven selections bias current demographic constraints toward the most extreme systems, potentially misrepresenting the diversity and evolution of the LRD population. Accounting for these selection effects is essential for interpreting LRDs and their role in early black hole growth.

\end{abstract}


\section{Introduction}

The advent of the JWST (\citealt{gardner_james_2023}) has revealed a previously unrecognized population of faint active galactic nuclei (AGNs) at high redshift, extending well below the luminosities probed by pre-JWST surveys (e.g., \citealt{harikane_jwstnirspec_2023, maiolino_jades_2023, pacucci_jwst_2023, scholtz_jades_2023, maiolino_jwst_2024}). Among the most striking of these discoveries is a class of sources now commonly referred to as {\it Little Red Dots} (LRDs; \citealt{matthee_little_2024}); these objects, now identified across multiple JWST extragalactic fields through Near Infrared Camera (NIRCam; \citealt{rieke_performance_2023}) surveys, are increasingly associated with a key phase of early black hole growth (\citealt{inayoshi_little_2025, pacucci_cosmic_2025}).

LRDs are defined by two key observational properties at rest-frame optical wavelengths: compact morphology and red colors. Their spectral energy distributions (SEDs), typically characterized by a “V-shaped” (in $F_{\lambda}$ vs. $\lambda$) continuum with blue rest-UV emission that steepens toward the rest-optical, cannot be easily reproduced by either star-forming galaxy or AGN templates. Notably, observations have shown that the minimum of this “V-shaped” feature is consistently observed near the Balmer limit in the rest frame (e.g., \citealt{furtak_high_2024, wang_rubies_2024, juodzbalis_jades_2024, setton_little_2024}). Recent studies, however, show that increasing continuum curvature and varying AGN–host contributions can broaden and partially wash out this feature, such that the apparent minimum shifts toward longer wavelengths (see discussion in \citealt{barro_cliff_2025}).

Interestingly, it has been also shown that a large fraction of photometrically selected LRDs show broad permitted lines (e.g., \citealt{hviding_rubies_2025}), despite their relatively modest bolometric luminosities and faint UV emission (e.g., \citealt{de_graaff_rubies_2025}). The estimated number densities of LRDs exceed those of UV-selected quasars by up to $1\text{--}2$ dex at comparable redshifts and absolute UV magnitudes ($M_{\rm UV}$; e.g., \citealt{kokorev_census_2024, kocevski_rise_2025, akins_cosmos-web_2025}), particularly at $z \gtrsim 4$, suggesting that obscured or partially obscured accretion may dominate the early growth of supermassive black holes. However, observations with the Mid-Infrared Instrument (MIRI; \citealt{wright_mid-infrared_2023}) show mixed results: many LRDs lack the strong mid-infrared emission expected from heavily dust-obscured AGN tori (e.g., \citealt{perez-gonzalez_what_2024}), while stacked MIRI analyses suggest the opposite (e.g., \citealt{delvecchio_active_2025}). At the same time, observations have shown that LRDs represent only a fraction of the overall galaxy population, suggesting that they correspond to a short-lived or highly specific evolutionary phase rather than a ubiquitous mode of galaxy growth (\citealt{inayoshi_little_2025}).

Since their initial discovery, the extragalactic community has made efforts to establish and refine robust photometric strategies to identify LRDs in JWST legacy surveys. Indeed, LRDs have now been identified using multiple, conceptually related selection strategies, including approaches based on fixed color criteria designed to isolate the characteristic “V-shaped” SED (e.g., \citealt{barro_extremely_2024, kokorev_census_2024, perez-gonzalez_what_2024, akins_cosmos-web_2025, rinaldi_not_2025, barro_comprehensive_2026}), as well as methods that infer rest-frame UV and optical continuum slopes using redshift-dependent filters (\citealt{kocevski_rise_2025}). These approaches have substantially expanded LRD samples, improved control over contamination from, e.g., galaxies with strong Balmer breaks or emission-line–dominated SEDs, and enabled systematic studies of LRD demographics across a broad redshift range, predominantly at $z \gtrsim 4$. 

By contrast, constraints at lower redshift remain limited: most of the current knowledge of the LRD population at $z \lesssim 4$ is derived from searches based either on ground-based facilities \citep{ma_counting_2025} or from {\it Euclid} \citep{bisigello_euclid_euclid_2025}\footnote{One must consider that these studies are (1) generally shallower than current JWST-based searches at $z\gtrsim4$, and (2) rely on facilities with markedly different capabilities (e.g., spatial resolution), which may introduce systematic biases in the inferred LRD demographics at $z<4$ and complicate direct comparisons with JWST results.}. These studies consistently point to a rapid decline in the number density of LRDs toward $z \lesssim 4$, in agreement with the limited number of JWST-based analyses that have extended LRD searches to lower redshifts (e.g., \citealt{ kocevski_rise_2025, barro_comprehensive_2026}). To date, the emerging picture is that LRDs appear rapidly at $z \approx 8$, peak in abundance at $z \approx 5\text{--}6$, and decline sharply toward $z \lesssim 4$, plausibly tracing evolving gas supply, obscuration geometry, and black hole fueling (\citealt{inayoshi_little_2025, pacucci_cosmic_2025}).

However, despite this rapid progress, a fundamental aspect unifying most recent LRD searches has received comparatively little scrutiny: by construction, photometric searches for LRDs require them to be extremely red at rest-frame optical wavelengths. Across the literature, different selection formalisms converge on this same requirement. For example, \citet{akins_cosmos-web_2025} and \citet{barro_extremely_2024} impose $\mathrm{F277W}\text{--}\mathrm{F444W} > 1.5$ mag, while \citet{kocevski_rise_2025} adopt a positive rest-frame optical slope ($\beta_{\rm opt} > 0$), which at $z\approx5$ corresponds to $\mathrm{F277W}\text{--}\mathrm{F444W}\gtrsim1$ mag, similar to the criteria adopted in, e.g., \citet{labbe_population_2023, greene_uncover_2024, perez-gonzalez_what_2024}. As a consequence, existing studies have systematically targeted only the most extreme red tail of this new population of objects.

Nonetheless, recent results increasingly suggest that this extreme subset does not capture the full phenomenology of LRDs (\citealt{hainline_investigation_2024, perez-gonzalez_little_2026}). Instead, LRDs appear intrinsically heterogeneous, spanning a broader range of optical colors and spectral properties than initially observed \citep{barro_cliff_2025, perez-gonzalez_little_2026}. This raises a natural question: {\it how many LRDs are missed in JWST legacy surveys because they fail the most extreme redness cuts?} More broadly, what new physical insight emerges when the LRD population is explored across a continuous range of optical redness, and how does the choice of color threshold reshape the observed demographics, number densities, and inferred redshift evolution of the population?

In this work, we present a census of photometrically selected LRDs in the \emph{JWST Advanced Deep Extragalactic Survey} (JADES) fields, exploiting the depth and homogeneous multiwavelength coverage of the latest JWST/NIRCam data release (DR) 5 (\citealt{johnson_jwst_2026, robertson_jwst_2026}). Building on recent selection strategies, we adopt deliberately relaxed color criteria to probe an extended range of optical redness rather than only the most extreme systems. This approach enables a statistically representative sample over $z\approx2\text{--}11$, while maintaining comparability with previous JWST studies. 

Our aim is to quantify how selection choices affect inferred demographics and to provide a uniform reference sample for future follow-up. The goals of this paper are threefold: (i) to measure the abundance and redshift evolution of LRDs using an intentionally more inclusive photometric selection; (ii) to evaluate how varying the redness threshold impacts number densities and luminosity distributions; and (iii) to place LRDs within the broader context of early black hole growth and AGN demographics.

Following \citet{labbe_population_2023} and \citet{barro_comprehensive_2026}, we define LRDs as compact systems that are red at rest-frame optical wavelengths and blue in the rest-frame UV. Under this broad definition, selected sources span a wide range of optical redness, from the most extreme systems isolated by stringent cuts \citep[e.g.,][]{kocevski_rise_2025, akins_cosmos-web_2025} to less extreme objects recovered with more inclusive criteria \citep[e.g.,][]{kokorev_census_2024, rinaldi_not_2025}. Albeit recent studies have begun distinguishing LRDs from their less optically red counterparts, the so-called Little Blue Dots (LBDs; e.g., \citealt{brazzini_little_2026}), based on differences in continuum slopes\footnote{\citet{brazzini_little_2026} define LRDs as sources with $\beta_{\rm opt} > 0$ and LBDs as sources with $\beta_{\rm opt} < 0$.} and excitation properties (see discussion in \citealt{brazzini_little_2026}), we do not subdivide our sample. Instead, we treat all selected sources as a single, operationally defined LRD population (i.e., compact and with a “V-shaped” SED; \citealt{labbe_population_2023, barro_extremely_2024}), consistent with the original photometric definition, leaving any physically meaningful distinction to future spectroscopic follow-up on statistically significant samples such as the one compiled in this work.

The paper is organized as follows. In Section 2, we describe the data sets adopted in this work. In Section 3, we review the selection criteria commonly used to identify LRDs, introduce the criteria adopted here, and discuss the parameter space they probe, together with the general properties of the resulting sample. In Section 4, we derive the UV and optical luminosity functions, and the redshift evolution of the LRD number density, comparing our results with those in the recent literature and examining how our intentionally more relaxed selection performs relative to the stricter cuts typically employed. In Section 5, we discuss the implications of our findings and summarize our main conclusions.

Throughout this paper, we consider a cosmology with $H_{0} = 70\; \rm km\;s^{-1}\;Mpc^{-1}$, $\Omega_{M} = 0.3$, and $\Omega_{\Lambda} =0.7$. All magnitudes are total and refer to the AB system \citep{oke_secondary_1983}. A \citet{kroupa_variation_2001} initial mass function (IMF) is assumed (0.1--100 M$_{\odot}$).

\section{Data and Photometry}

In this work, we adopt the public JADES DR5 imaging and photometric catalogs in the GOODS-South (GOODS-S) and GOODS-North (GOODS-N) fields (\citealt{johnson_jwst_2026, robertson_jwst_2026}). The DR5 dataset combines $\approx$1250 hours of JWST observations with extensive ancillary data, providing uniformly processed, multiwavelength photometry across 35 space-based bands from the optical to the mid-infrared.

The DR5 imaging consists of fully reduced, astrometrically aligned, and mosaicked JWST (NIRCam and MIRI) and archival HST imaging, and is described in detail by \citet{johnson_jwst_2026}. In particular, it includes HST/ACS imaging (F435W, F606W, F775W, F814W, F850LP) and HST/WFC3 imaging (F105W, F125W, F140W, F160W; \citealt{whitaker_hubble_2019}), together with JWST/NIRCam coverage spanning both wide and medium bands (F070W, F090W, F115W, F150W, F162M, F182M, F200W, F210M, F250M, F277W, F300M, F335M, F356W, F410M, F430M, F444W, F460M, F480M)\footnote{GOODS-N lacks F250M and F480M; see Table~1 of \citet{robertson_jwst_2026} for a detailed summary of the filter coverage and depths.}, assembled from multiple programs (e.g., FRESCO, JEMS, MIDIS, NGDEEP, CONGRESS, PANORAMIC; \citealt{oesch_jwst_2023, perez-gonzalez_life_2023, williams_jems_2023, bagley_next_2024, lin_luminosity_2025, williams_panoramic_2025}), including JADES itself (\citealt{eisenstein_overview_2023}). On the MIRI side, the JADES DR5 incorporates JWST/MIRI imaging from F560W to F2550W, combining JADES parallel observations (\citealt{alberts_jwst_2026}) with the SMILES mosaics (\citealt{alberts_smiles_2024}). The NIRCam data reach typical depths of $\gtrsim 29\text{--}30$~mag ($5\sigma$, point sources) over areas of $\gtrsim 200~\mathrm{arcmin}^2$. 

In the JADES DR5 photometric catalog, sources are detected in signal-to-noise--weighted stacks of the long-wavelength NIRCam imaging, deblended using higher-resolution short-wavelength data, and characterized through a consistent framework that includes circular aperture photometry, ellipsoidal Kron photometry, and curve-of-growth measurements in every band. Photometric uncertainties explicitly account for correlated noise across the heterogeneous mosaics, and photometric redshifts are derived in a uniform manner using template-based fitting applied to both small-aperture and common point spread function (PSF) photometry (see also \citealt{hainline_jwst_2026}). This combination of depth, wavelength coverage, and homogeneous processing provides a robust foundation for constructing reliable SEDs and for statistical analyses of galaxy populations across cosmic time, which we adopt throughout this work. For a full description of the methodology, we refer the reader to \citet{robertson_jwst_2026}.

\section{Sample Selection and Source Characterization}

The ultra-deep JWST/NIRCam imaging from the JADES survey provides an ideal dataset for a photometric search of LRDs. The combined HST and JWST observations deliver continuous wavelength coverage from $\approx 0.4$ to $5\,\mu$m across a dense set of broad and medium bands, reaching typical depths of $\gtrsim 29\text{--}30$~mag ($5\sigma$, point sources; see Table~1 in \citealt{robertson_jwst_2026}) over areas of $\gtrsim 200~\mathrm{arcmin}^2$.

\subsection{Photometric redshifts}

We adopt the “CIRC1” photometric redshifts from the public JADES DR5 catalog, derived as described in \citet{robertson_jwst_2026}. These redshifts are obtained by fitting the $0.2''$-diameter circular-aperture JADES photometry using the Python implementation of \texttt{EAZY} \citep{brammer_eazy_2008}. We employ the \texttt{EAZY} template set introduced by \citet{hainline_cosmos_2024}, which includes templates with red optical slopes suitable for modeling LRD SEDs. Photometric redshifts are determined at the minimum $\chi^2$, allowing \texttt{EAZY} to fit linear combinations of all templates. Comparisons with available spectroscopic redshifts indicate that these photometric redshifts are generally accurate (\citealt{rieke_jades_2023, robertson_jwst_2026}), although we caution that LRDs can be faint at $\lambda<2\,\mu$m, where low signal-to-noise measurements of the Lyman-$\alpha$ break may lead to increased uncertainties or biases toward higher redshifts (\citealt{kocevski_hidden_2023, hainline_investigation_2024}). The exceptional depth of the JADES imaging, together with the inclusion of both wide and medium-band filters, helps to mitigate these effects.

\subsection{The search for Little Red Dots in the JADES fields}

There now exist multiple approaches for selecting LRDs, albeit with some differences among them. The literature has demonstrated the effectiveness of these methods, and the associated criteria have been rapidly refined by the community over a relatively short timescale. At present, at least two main photometric strategies are commonly adopted: one based on color selections (e.g., \citealt{kokorev_census_2024, rinaldi_not_2025, hainline_investigation_2024, barro_comprehensive_2026, akins_cosmos-web_2025}), and another relying on continuum-slope measurements in the rest-frame UV and optical (e.g., \citealt{lin_discovery_2024, kocevski_rise_2025}). 

In both cases, these photometric criteria are typically combined with morphological constraints (most notably compactness in F444W) and with additional requirements designed to suppress contamination from interlopers that can mimic LRD colors, such as brown dwarfs. In the following, we first summarize for the reader the commonly adopted criteria used to photometrically select LRDs, and then detail the specific thresholds and parameter choices adopted in this work.

\subsubsection{Color selection criteria} 

Building on the traditional selection of extremely red objects (EROs) through simple color cuts (e.g., $R-K$; \citealt{elston_deep_1988, elston_observations_1989}), JWST now extends this approach to fainter limits and longer wavelengths using NIRCam colors. Indeed, the strategy of targeting extremely red sources was initially adopted to identify LRDs (e.g., \citealt{labbe_population_2023}), and has since been maintained through the use of stringent color thresholds to ensure robust candidate selection and minimize contamination.

For example, \citet{akins_cosmos-web_2025} applied a stringent cut of $\rm F277W\text{--}F444W > 1.5$ mag (similar to \citealt{barro_extremely_2024})\footnote{For instance, \citet{greene_uncover_2024} adopted a threshold of 1~mag.}, while \citet{kocevski_rise_2025} relied on continuum-slope criteria ($\beta_{\rm opt} > 0$ and $\beta_{\rm UV} < -0.37$), which at $z\approx5$ effectively correspond to selecting sources with $\rm F277W\text{--}F444W \gtrsim 1.0$ mag and $\rm F150W\text{--}F277W < 0.5$ mag\footnote{The key difference is that fixed color cuts do not explicitly account for redshift evolution, whereas the slope-based approach estimates continuum slopes using multiple filters as a function of redshift, providing a more flexible, redshift-aware selection.}. These choices are primarily pragmatic rather than physically motivated: adopting redder color thresholds efficiently reduces contamination from blue galaxies with shallow stellar continua whose NIRCam fluxes can be artificially boosted by extremely high equivalent width (EW) emission lines (see \citealt{rinaldi_midis_2023} for the impact on photometry and \citealt{boyett_extreme_2024} for the spectroscopic evidence), an effect that becomes increasingly important for NIRCam-based selections at $z\gtrsim6\text{--}7$ (see discussion in \citealt{hainline_investigation_2024}).

However, as recently discussed by \citet{zhang_abundant_2025}, selections tuned to the most extreme colors or continuum slopes capture only a subset of the (perhaps) broader population (e.g., \citealt{perez-gonzalez_little_2026}). While such criteria efficiently identify the reddest and most conspicuous photometrically selected LRDs, thus making them well suited for initial discovery and contamination control (e.g., \citealt{hainline_investigation_2024}), they systematically miss sources that remain optically red but do not satisfy the most extreme thresholds. As a result, the classic selections preferentially probe only the reddest end of the LRD distribution, biasing samples against (perhaps) physically related but moderately red systems.

\subsubsection{Compactness} 

Compactness is a another key parameter in the selection of LRDs (see also results in \citealt{hviding_rubies_2025}). Previous studies have shown that these sources are typically extremely compact at rest-frame optical wavelengths (e.g., \citealt{baggen_small_2024}), while often exhibiting more complex or extended morphologies in the rest-frame UV (\citealt{rinaldi_not_2025, chen_host_2025}). This apparent diversity is likely driven, at least in part, by the substantial differences in spatial resolution and PSF between the NIRCam short- and long-wavelength channels\footnote{Over the redshift range $z\approx4\text{--}8$ (where LRDs are mostly studied), the NIRCam PSF corresponds to physical scales of $\approx100\text{--}350$~pc in the short-wavelength channel and $\approx700\text{--}1000$~pc in the long-wavelength channel. Consequently, while rest-frame UV imaging probes structure on a few hundred parsec scales, rest-frame optical observations smooth emission over nearly kiloparsec scales. This difference in effective resolution naturally contributes to the observed UV–optical morphological differences (\citealt{rinaldi_not_2025})}. 

Different methods have therefore been adopted in the recent literature to select compact sources. The most commonly used approach relies on flux ratios measured in the F444W band within apertures of different sizes (e.g., \citealt{kokorev_census_2024, rinaldi_not_2025, akins_cosmos-web_2025}). An alternative method employs a magnitude-dependent size cut based on the half-light radius measured in F444W. In this case, the stellar locus is identified in the $m_{\rm F444W}$–$r_{\rm h}$ plane, and compact sources are selected relative to this locus, with the allowed tolerance increasing toward fainter magnitudes to account for larger measurement scatter (see \citealt{kocevski_rise_2025}).

\subsubsection{Brown Dwarf removal} Finally, empirical tests have shown that selections based on compactness and red colors alone do not yield a fully pure LRD sample, as they can be contaminated by sources that mimic LRD colors (especially in the blue part), most notably low-temperature (e.g. spectral class T and Y) brown dwarfs. Several strategies have been proposed in the literature to photometrically identify and exclude such interlopers. Brown dwarfs, which are unresolved, have atmospheric molecular absorption which leads to a blue observed slope at $1\text{--}3\,\mu$m and a bright peak at $4\text{--}5\,\mu$m which can mimic the color of LRDs. 

One commonly adopted approach for excluding brown dwarfs from LRD samples relies on dedicated color cuts motivated by spectroscopic results from the UNCOVER survey (\citealt{bezanson_jwst_2024}), which demonstrated that brown dwarf contaminants can be efficiently identified and removed (namely $\rm F115W\text{--}F200W >-0.5$~mag; \citealt{greene_uncover_2024}). In addition, \citet{hainline_jades_2025} introduced an alternative color-based criterion to isolate the brown dwarf locus in NIRCam color–color space (see their Section~3 and Figure~1). 

\subsubsection{Criteria adopted in this work}

In this work, we adopt a selection strategy that balances inclusiveness with robustness by combining relaxed color-based criteria with stricter requirements on compactness and contamination rejection. This approach is designed to avoid biasing the sample toward only the most extremely red LRDs (e.g., $\rm F277W\text{--}F444W>1.5$~mag), yielding a more inclusive population that is better suited for future follow-up observations and for probing the physical nature of these sources as well as their (perhaps) diversity. Building on the selection proposed by \citet{rinaldi_not_2025}, we relax the color criteria from $\rm F150W\text{--}F200W < 0.8$~mag to $\rm F150W\text{--}F200W < 1.0$~mag, and from $\rm F277W\text{--}F444W > 0.7$~mag to $\rm F277W\text{--}F444W > 0.5$~mag\footnote{Although our primary aim is to broaden the rest-frame optical selection, we slightly relax the UV cut by 0.2\,mag with respect to the original criterion adopted in, e.g., \citet{rinaldi_not_2025, kokorev_census_2024}. As LRDs are typically faint in the UV, colors are more susceptible to photometric scatter. The relaxed $<1.0$\,mag cut preserves the blue UV requirement while minimally affecting the final sample ($\lesssim10\%$ increase).}. Typical emission-line galaxies (at $z\gtrsim3$) with H$\beta$+[O\,\textsc{iii}]$\lambda\lambda4959,5007$ and H$\alpha$ exhibit rest-frame EWs of $\approx600$--$800$\,\AA\, (e.g., \citealt{boyett_extreme_2024}). For the adopted NIRCam filters, this corresponds to an average photometric excess of $\approx0.5$\,mag over $z\approx4$--$8$ when adopting the formula from \citet{marmol-queralto_evolution_2016}. This expectation motivates adopting a relaxed red optical threshold of $\rm F277W\text{--}F444W>0.5$~mag.

Moreover, \citet{zhang_abundant_2025} showed that the original criteria from \citet{rinaldi_not_2025} are substantially more inclusive than other commonly adopted selections (e.g., from \citealt{barro_extremely_2024} or \citealt{kocevski_rise_2025}). For instance, the criteria from   \citet{barro_extremely_2024} would recover only $\approx$15–23\% of their final LRD sample (see  Figure~8 in \citealt{zhang_abundant_2025}), whereas the \citet{rinaldi_not_2025} selection captures a significantly larger fraction (up to $\approx60\%$). However, broader criteria may also capture strong emission-line galaxies, which can be efficiently identified through visual inspection. Conversely, adopting very stringent color cuts (e.g., \citealt{akins_cosmos-web_2025}) restricts the selection to the most extreme region of color–color space and excludes sources occupying the less extreme portion of the LRD-like locus\footnote{We stress that it is not known a priori whether these excluded sources belong to the same underlying population of the one studied in, e.g., \citet{akins_cosmos-web_2025}; however, it is equally unclear whether the most stringent cuts isolate the full phenomenology of LRDs.}. 

\begin{figure*}[ht!]
    \centering
    \includegraphics[width=1.0\linewidth]{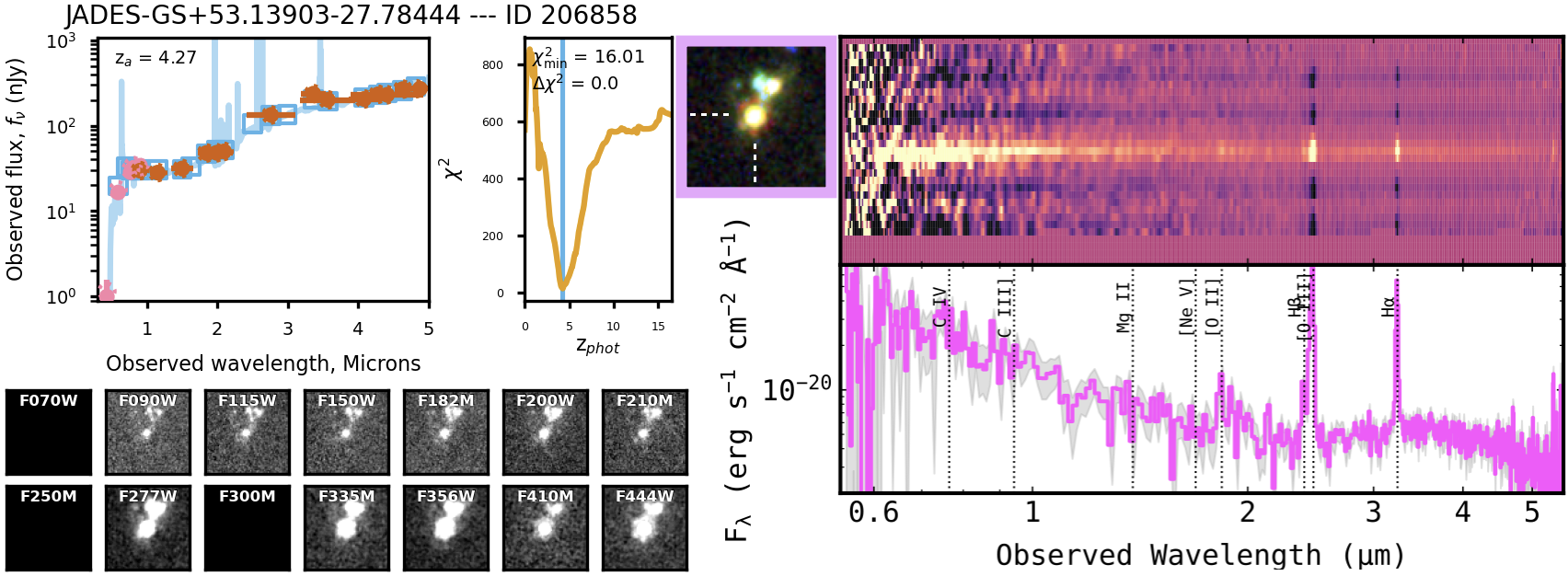}
    \caption{Example of an LRD candidate at $z_{\rm spec}\approx3.94$ ($z_{\rm phot}=4.27$, well within the expected scatter and not an outlier) that would be missed by standard color- or slope-based selections, with $\rm F277W\text{--}F444W\approx0.52$ mag and $\beta_{\rm opt}\approx-0.33$, placing it outside commonly adopted LRD criteria. {\bf Left panel:} postage stamps and best-fit SED from {\sc eazy}, showing the characteristic flat UV and rising optical continuum (in $F_{\nu}$); the source exhibits complex UV morphology and is surrounded by blue companions at the same $z_{\rm spec}$, as seen in other LRDs (see \citealt{rinaldi_not_2025}), with redshifts confirmed by NIRSpec and FRESCO (\citealt{curtis-lake_jades_2025, scholtz_jades_2025, oesch_jwst_2023}). {\bf Right panel:} the NIRSpec/PRISM spectrum reveals a prominent UV component together with the characteristic curved optical/NIR continuum of LRDs (in $F_{\lambda}$).}
    \label{fig:missed_LRD}
\end{figure*}

An illustrative example is shown in Figure~\ref{fig:missed_LRD}. The source has $\rm F277W\text{--}F444W\approx0.52$\,mag and $\beta_{\rm opt}\approx-0.33$ (following \citealt{kocevski_rise_2025}), placing it outside most commonly adopted selection criteria. Nevertheless, it exhibits the curved SED at optical/near-infrared wavelengths and strong UV component typical of LRDs \citep{barro_cliff_2025}, along with a broad-line component\footnote{This source has also medium resolution grating observations (PID 1180) with a broad H$\alpha$ component.} with $\log_{10}(M_{\bullet}/M_{\odot}) = 6.86^{+0.15}_{-0.21}$ \citep{juodzbalis_jades_2025}. Although classified as a Type~1 AGN by \citet{juodzbalis_jades_2025}, its pronounced SED curvature is not typical of classical Type~1 AGNs; accordingly, only $\approx30\%$ (11/34) of their sample overlaps with our LRD sample.

In addition to the color selection, we apply a stricter compactness requirement motivated by recent results showing that LRD surface-brightness profiles are statistically indistinguishable from the F444W point-spread function (PSF; see, e.g., \citealt{whalen_limitations_2025}). Rather than adopting the commonly used F444W aperture--flux ratio as a size proxy\footnote{We note that different studies use different aperture--flux ratios.}, we rely on direct F444W size measurements from the JADES DR5 morphological catalog (\citealt{carreira_jwst_2026}), where galaxy sizes are derived using \textsc{pysersic} \citep{pasha_pysersic_2023}. We retain only systems that are consistent with being unresolved at the F444W resolution\footnote{This choice is conservative, as it may exclude genuinely extended LRDs; see discussion in \citet{billand_investigating_2025}.}, requiring $R_{\mathrm{eff}} \lesssim 0.06$\arcsec.

\vspace{3mm}

Finally, we apply brown dwarf rejection criteria following \citet{greene_uncover_2024} ($\rm F115W\text{--}F200W>-0.5$~mag), together with the additional color cuts introduced by \citet{hainline_jades_2025}. In the latter case, sources consistent with the brown dwarf locus are excluded by requiring $\rm (F277W\text{--}F444W) < 0.4\,(F115W\text{--}F277W) + 1.9$ and $\rm (F115W\text{--}F277W) > -0.5$. The latter cuts primarily remove sources lying near the boundary between the LRD and brown dwarf loci, indicating that the initial rejection based on \citet{greene_uncover_2024} is already largely effective. Cross-matching the excluded objects with available catalogs the Near-Infrared Spectrograph (NIRSpec; \citealt{ferruit_near-infrared_2022, jakobsen_near-infrared_2022}) in the JADES field reveals (\citealt{curtis-lake_jades_2025, scholtz_jades_2025}) two sources with $z_{\rm spec}\approx7$, suggesting that some objects near the locus boundary may not be intrinsically brown dwarfs. We therefore regard the Hainline et al. cuts as conservative, as they may exclude a small number of genuine LRDs close to the separation between the two populations. Nevertheless, we adopt them to maximize the purity of the final sample. Our final selection also requires a signal-to-noise ratio $\mathrm{SNR} > 3$ in the F444W band.

\subsubsection{The final sample}
Our selection criteria yield 598 objects in GOODS-S (out of 304,366 sources) and 218 objects in GOODS-N (out of 181,144 sources) over a total area of $349.6~\mathrm{arcmin}^2$ (comprising  $210.44~\mathrm{arcmin}^2$ in GOODS-S and $139.16~\mathrm{arcmin}^2$ in GOODS-N). All of these sources were visually inspected (see next paragraphs), resulting in a final sample of 220 objects in GOODS-S and 101 in GOODS-N, for a total of 321 sources (roughly one object per arcmin$^{2}$) constituting the main sample analyzed in this work. 

As a sanity check, motivated by recent results showing that photometrically selected LRDs can also be contaminated by quiescent galaxies (in addition to brown dwarfs and strong-line emitters), we cross-matched our sample with the catalog of (photometrically selected) quiescent galaxies in the JADES fields presented by \citet{baker_exploring_2025}  \citep[using the criteria of][]{baker_abundance_2025}. We find three matches. Among these three sources, two show clear photometric excess in the medium bands at the wavelengths expected for H$\beta$+[O\,{\sc iii}] and H$\alpha$ (IDs 152330 and 226386; JADES-GS+53.1027-27.7444 and JADES-GS+53.1912-27.8357), indicating the presence of emission lines and therefore disfavoring a quiescent interpretation \citep[although an AGN within a quiescent galaxy remain a potential explanation; e.g.,][]{carnall_jwst_2024,kokorev_silencing_2024, baker_abundance_2025}. The third source is excluded from our sample, as we cannot draw firm conclusions on its nature. For JADES-GS+53.1027$-$27.7444, a NIRCam/WFSS spectrum from FRESCO is available and reveals a broad H$\alpha$ component with a full width at half maximum (FWHM) of $2068.78 \pm 40.82$~km s$^{-1}$, which was not reported in previous analyses based on FRESCO data such as \citet{matthee_little_2024}.

From our visual inspection, two main behaviors emerge. First, selecting LRDs at $z \gtrsim 6\text{--}7$ based solely on NIRCam photometry is intrinsically challenging. At $z\approx6\text{--}7$, both H$\beta$ and [O\,{\sc iii}]$\lambda\lambda4959,5007$ as well as H$\alpha$ fall at observed wavelengths $\gtrsim4\,\mu$m, potentially boosting the $\mathrm{F277W\text{--}F444W}$ color and mimicking a rising continuum if one were to rely on simple color cuts (or even on optical slopes; \citealt{kocevski_rise_2025}). In this regime, a genuinely rising continuum can often be distinguished from line-driven photometric excess through visual inspection of the SED, especially when dense filter coverage (also including medium bands)  such as that provided by JADES is available (although not across the full area). In contrast, at $z\gtrsim7$, H$\alpha$ shifts beyond the NIRCam wavelength coverage (\citealt{rinaldi_midis_2023}), leaving H$\beta$ and [O\,{\sc iii}]$\lambda\lambda4959,5007$  as the dominant contributors to flux enhancements in the reddest NIRCam bands; in this case, the combined H$\beta$+[O\,{\sc iii}]$\lambda\lambda4959,5007$  emission can closely mimic a genuinely rising SED, with the resulting degeneracy becoming increasingly severe toward higher redshift, thus making visual inspection extremely important.

Indeed, during our visual inspection, strong line emitters and sources with ambiguous, apparently rising continua constituted the dominant source of contamination, as also noted by \citet{hainline_investigation_2024}, particularly at $z \gtrsim 6\text{--}7$, accounting for up to $\approx80\%$ of the rejected sample. However, visual inspection alone does not provide a direct constraint on the intrinsic continuum shape, making it difficult to robustly assess the true emission-line contribution and disentangle line-driven excess from a genuinely rising continuum. To test for possible biases arising from this degeneracy (especially at $z>7$), we therefore performed an additional consistency check on the surviving sources using \textsc{bagpipes} \citep{carnall_vandels_2019}, adopting an approach similar to that of \citet{kokorev_census_2024}. 

\begin{figure}[ht!]
    \centering
    \includegraphics[width=1.0\linewidth]{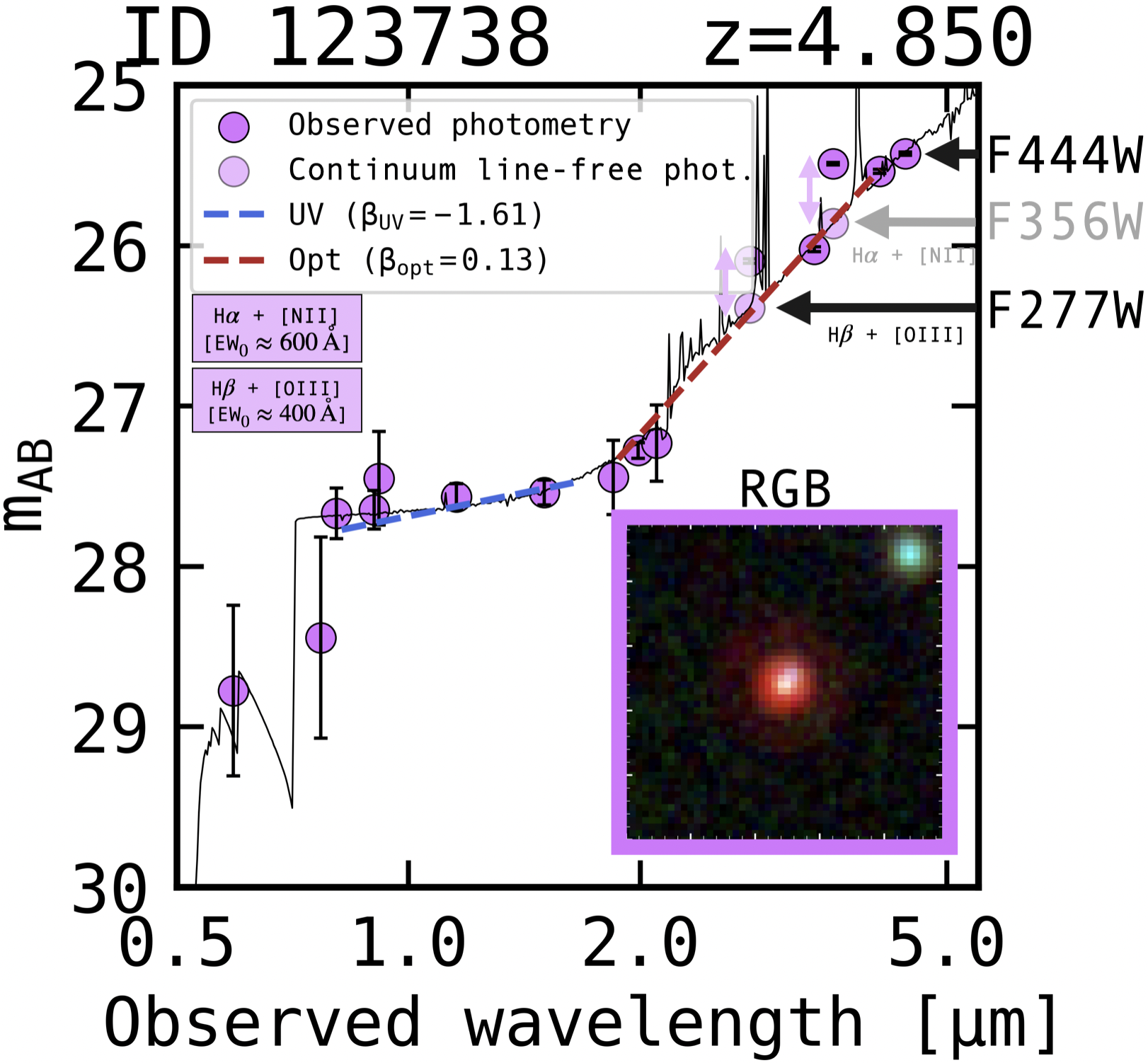}
    \caption{Example of a photometrically selected LRD in which the F277W band is contaminated by H$\beta$ + [O\,{\sc iii}]$\lambda\lambda4959,5007$, while H$\alpha$ contributes to the flux in F356W (not used in our selection). The rest-frame equivalent widths are $\rm EW_{0}\approx400\text{--}600$~\AA, leading to a photometric excess up to $\approx0.4$~mag in the affected bands. In this case, the observed photometry yields $\rm F277W-F444W\approx0.67$~mag, whereas correcting for line emission would imply a redder color of $\approx1$~mag. Importantly, the presence of emission lines does not bias our final selection. Interestingly, if the line-free F277W flux is used to estimate the optical slope, the source would satisfy $\beta_{\rm opt}>0$, while otherwise it would be $\beta_{\rm opt}<0$. Remarkably, although the source does not meet the most extreme selection thresholds (e.g., \citealt{barro_extremely_2024, kocevski_rise_2025, akins_cosmos-web_2025}), its SED still exhibits a steeply rising trend toward the optical, consistent with the characteristic behavior of LRDs.}
    \label{fig:test_EW}
\end{figure}

In this test, we explored a set of templates allowing for very strong nebular emission as recently observed with JWST (e.g., \citealt{boyett_extreme_2024}). This modeling explicitly accounts for emission-line contributions and enables us to recover continuum-only (line-free) photometry by correcting the broadband fluxes for the fitted line emission. We then re-applied the same color-based selection criteria to the continuum-only photometry for our visually inspected sample. We show an example of such application in Figure \ref{fig:test_EW}. 

This exercise recovers essentially all of the originally selected sources that survived our visual inspection, with the exception of a very small number of objects ($\lesssim10\%$), primarily at $z\gtrsim7$, where strong H$\beta$ and [O\,{\sc iii}]$\lambda\lambda4959,5007$ emission can significantly contaminate photometry at $\lambda_{\rm obs}>4\,\mu$m. After correcting the photometry for line boosting, the NIRCam data alone do not provide clear evidence for a genuinely rising continuum in these few cases. Given this ambiguity and the associated uncertainties, we conservatively exclude these sources from the final sample. A caveat of this approach is its reliance on photometric redshifts; however, in the JADES fields this limitation is largely mitigated by the wide wavelength baseline provided by the combined broad- and medium-band coverage, which yields robust photometric redshifts.

Second, by requiring sources to be unresolved in the F444W band, we mitigate biases toward extended systems (e.g., see Figure~1 of \citealt{akins_cosmos-web_2025}) and reduce the number of objects requiring visual inspection by up to $\approx30\%$ compared to a flux-ratio selection.

\vspace{2mm}

\begin{figure*}[ht!]
    \centering
    \includegraphics[width = 0.98 \textwidth, height = 0.4 \textheight]{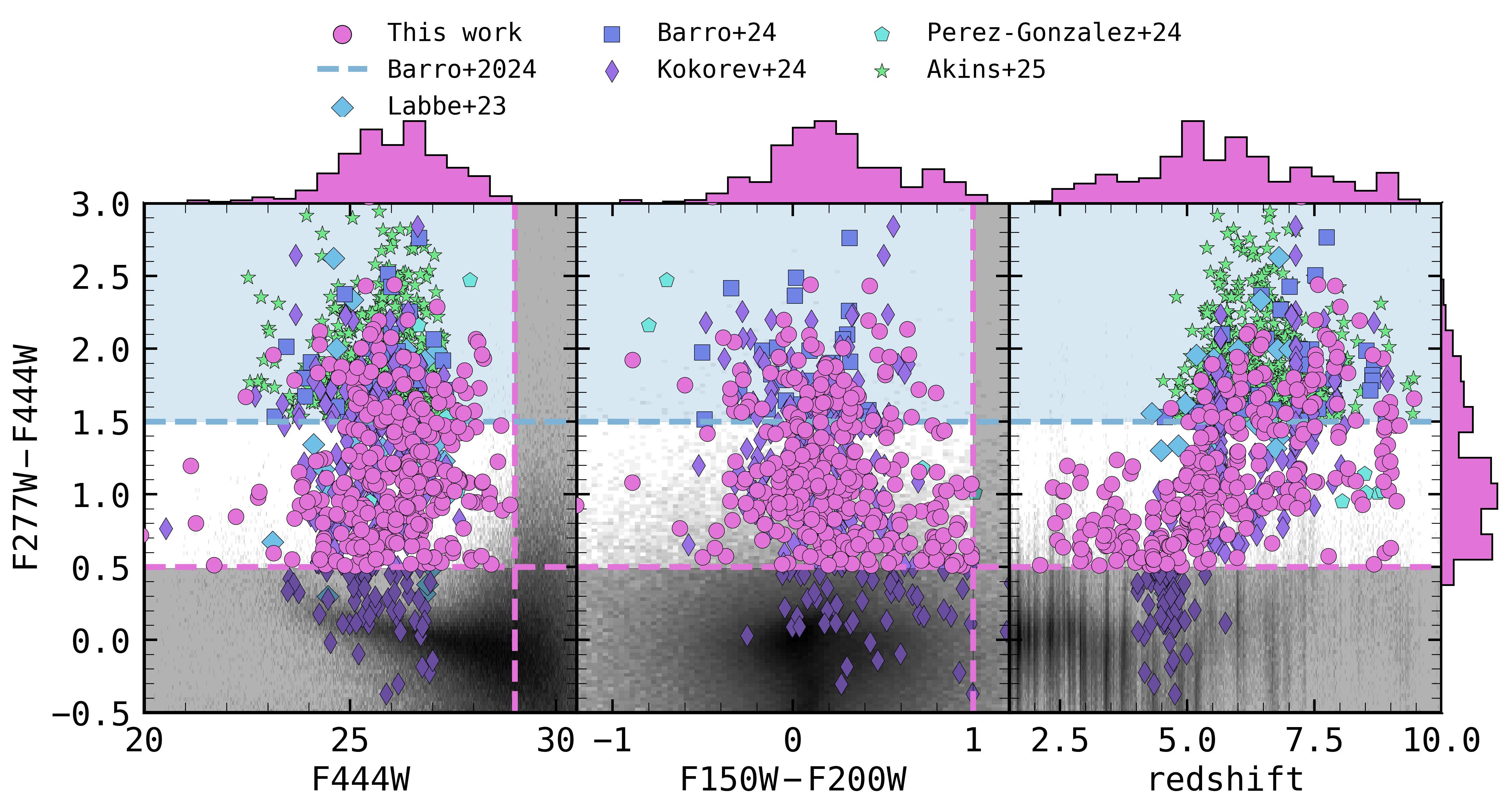}
    \caption{The photometrically selected LRD sample in the GOODS-N and GOODS-S fields, alongside other recent literature samples: \citet{labbe_population_2023, akins_cosmos-web_2025, barro_extremely_2024, kokorev_census_2024, perez-gonzalez_what_2024}. We also show that, if one were to apply the color cuts adopted in, e.g., \citet{barro_extremely_2024}, one would end up selecting only a subset of the population, with the majority of sources excluded. For each plotted property, we also show the corresponding histogram.} 
    \label{fig:color_cuts_vs_literature}
\end{figure*}

In Figure~\ref{fig:color_cuts_vs_literature}, we compare our sample with some of the LRD samples from the recent literature. Interestingly, two trends emerge. First, the distribution of $\rm F277W\text{--}F444W$ colors (right-hand histogram) shows that sources above the commonly adopted thresholds of $\approx1\text{--}1.5$~mag trace only the extreme red tail of the population, representing only a minority ($\lesssim25\%$) of the full LRD population (as also recently shown by \citealt{perez-gonzalez_little_2026} where the most extreme LRDs represent just a minor fraction of their total sample). This implies that the bulk of LRDs occupies a region of parameter space that has been less explored than the most extreme regime, yet may be critical for fully understanding the nature of this population. Recently, these less red LRDs (specifically with $\beta_{\rm opt}<0$) have been referred to as LBDs (e.g., \citealt{brazzini_little_2026}). However, detailed spectroscopic analyses, albeit on very small samples, suggest they likely trace the same underlying population, possibly observed at different evolutionary stages or with varying geometry or gas fractions (e.g., \citealt{asada_origins_2026, brazzini_little_2026, perez-gonzalez_little_2026}).

Second, the redshift distribution reproduces the well-established evolutionary trend of LRDs with cosmic time, with the number of sources peaking at $z\approx5\text{--}6$ (e.g., \citealt{pacucci_cosmic_2025, inayoshi_little_2025}) and then declining sharply toward lower redshifts, becoming rare at $z\lesssim4$ (e.g., \citealt{kokorev_census_2024, kocevski_rise_2025, bisigello_euclid_euclid_2025, ma_counting_2025}). 

We also observe a decline in the number of LRDs at $z\gtrsim7\text{--}8$, which may reflect intrinsic physical evolution (\citealt{inayoshi_little_2025}) or observational effects such as cosmological surface-brightness dimming (\citealt{billand_investigating_2025, pacucci_cosmic_2025}). However, one must also consider that a key limitation of current JWST surveys is that they rely predominantly on NIRCam, which probes progressively bluer rest-frame wavelengths at high redshift and becomes less effective at identifying LRDs. Consistently, many sources rejected by our selection lie at $z\gtrsim6\text{--}7$, where NIRCam alone struggles to distinguish genuinely rising continua from line-driven photometric excess.

At high redshifts, MIRI provides direct access to the rest-frame optical emission of LRDs, enabling a more robust identification of these systems at early epochs. Indeed, as also outlined in \citet{barro_cliff_2025}, the source {\it Virgil} provides one of the clearest demonstrations of the MIRI capability to date (\citealt{iani_midis_2024, rinaldi_deciphering_2025}); other examples include {\it Cerberus} (\citealt{perez-gonzalez_nircam-dark_2024}), which can be classified as an LRD candidate at $z\approx14$, and a recent LRD candidate at $z\approx10$ reported by \citet{tanaka_discovery_2025} in COSMOS-Web.  At present, however, MIRI observations remain severely limited, not only in surveyed area but, more critically, in depth. As a sanity check, we cross-matched our sample with the MIRI Legacy Extragalactic Survey (SMILES) catalog (\citealt{alberts_smiles_2024, rieke_smiles_2024}) in GOODS-S, which can be considered as representative of current wide-area MIRI surveys and their typical depths (e.g., COSMOS-Web and EGS; \citealt{backhaus_mega_2025, finkelstein_cosmic_2025, harish_cosmos-web_2025}), and found 31 matches\footnote{In at least one MIRI band.} out of 50 photometrically selected LRDs within the SMILES footprint; the remaining sources, with $\mathrm{F444W}\approx27\text{--}29$ mag, fall below the MIRI detection limits (typically $\approx25\text{--}25.5$ mag in the bluest bands), indicating that current shallow MIRI data recover only the brightest LRDs. Moreover, the redshift distribution of the MIRI-detected LRDs shows that $\approx84\%$ lie at $z\lesssim7$, highlighting that such observations preferentially trace lower-redshift LRDs, primarily around $z\approx4\text{--}5$ in our case\footnote{These are typically bright in NIRCam/F444W ($\lesssim26$~mag).}.

\subsection{The impact of specific selection criteria on LRD samples}

The selection criteria adopted in this work rely on fixed color cuts drawn from the recent literature and represent a widely used empirical approach for identifying LRDs. By construction, however, such selections are insensitive to the redshift: the same filter combinations probe different regions of the SED as redshift changes. As a result, fixed filters effectively select red objects regardless of which part of the continuum is sampled, and their ability to isolate genuine LRDs decreases outside the redshift range where the characteristic LRD SED is bracketed by the chosen bands.
For instance, our primary criteria do not select the Rosetta Stone at $z\approx2.26$ reported by \citet{juodzbalis_jades_2024}, as at this redshift the $\mathrm{F277W}\text{--}\mathrm{F444W}$ color probes the rest-frame near-infrared continuum and has a relatively modest value of $\approx0.32$ mag.

More sophisticated approaches have therefore been proposed, for example adapting the filter combinations with redshift so that similar rest-frame regions are consistently probed (\citealt{kocevski_rise_2025}). While physically motivated, such methods depend strongly on the available filter set, the goodness of the photometric redshifts, and on assumptions about which spectral regions should define the LRD selection. The latter is particularly relevant given recent suggestions that LRDs may span a continuum of SED shapes (e.g., \citealt{barro_cliff_2025, perez-gonzalez_little_2026}), ranging from strongly curved optical/NIR spectra to nearly power-law continua, potentially biasing selections against systems in the latter regime.

To partially mitigate these effects, we explored complementary color criteria to identify candidates at $z\lesssim4$ using only JWST photometry in the JADES fields, while the primary selection is optimized for $z\gtrsim4$ objects. To do so, we explored a complementary approach anchored to F090W. By separating the rest-frame UV and optical/NIR emission with $\rm F090W\text{--}F115W$ (or $\rm F090W\text{--}F150W$)\footnote{These combinations can be used interchangeably, as the final number of selected LRDs differs by less than 10\%.} and $\rm F200W\text{--}F444W$\footnote{With this last configuration, we successfully recover the Rosetta Stone.}, and enforcing the same compactness criterion ($R_{\rm eff}\lesssim0.06$\arcsec), we visually inspected $\approx300$ additional sources and identified 91 new photometric LRD candidates, predominantly at $z<5$ ($\approx80\%$).

Finally, we also explored multiple filter combinations to assess whether different band pairs are better suited above and below $z\approx4$. We identify the combinations that minimize the number of inspected sources while maximizing the number of confirmed LRDs: $\rm F150W\text{--}F200W$ and $\rm F277W\text{--}F444W$ for $z>4$, and $\rm F090W\text{--}F115W$ (or $\rm F090W\text{--}F150W$) together with $\rm F200W\text{--}F444W$ for $z<4$.

Overall, although more sophisticated selections (e.g., band-shifting approaches; \citealt{kocevski_rise_2025}) can better account for the redshift dependence of the SED, simple color cuts remain a practical first-order method for identifying LRD candidates in large photometric datasets, particularly when dense filter coverage is not available.

\subsection{Spectroscopically confirmed Little Red Dots in the JADES fields}

Our final sample comprises 321 sources across GOODS-S and GOODS-N selected using our primary criteria, together with an additional 91 sources identified using an alternative selection tailored to lower-redshift LRDs, for a total of 412 candidates (some examples are shown in Figure \ref{fig:example_LRDs} in Appendix \ref{appendix_example}). In the following, we will briefly explore the photometrically selected LRDs with confirmed spectroscopic redshifts.

\subsubsection{Cross-match with existing spectroscopic catalogs}

We cross-matched the photometrically selected sample with all available spectroscopic redshifts in the JADES fields, obtaining 121 objects with spectroscopy from NIRCam/grism (FRESCO and CONGRESS; \citealt{oesch_jwst_2023, lin_luminosity_2025}), MUSE (\citealt{urrutia_muse-wide_2019, bacon_muse_2023}), and NIRSpec (namely, JADES DR4 and the DAWN JWST Archive; \citealt{curtis-lake_jades_2025, scholtz_jades_2025, heintz_strong_2024}). 

In particular, we find that 41 photometrically selected LRDs have NIRSpec data, including 34 PRISM observations and 7 higher-resolution spectra. This comparison validates both the photometric redshifts and the robustness of the selection, yielding 11 outliers\footnote{Where an outlier is defined as follows: $|z_{\rm phot}-z_{\rm spec}|/(1+z_{\rm spec})>0.15$.} (10\%) with a scatter of $\sigma_{\rm NMAD}=0.04$ and a small bias of $\langle \delta z \rangle = +0.03$. An additional 10 sources have NIRSpec/PRISM spectroscopy from the DIVER program (PID: 8018; PI: X. Lin; \citealt{lin_diver_2025}), independently confirming their LRD nature. Only one object is an outlier at $z_{\rm spec}\approx2.04$ ($z_{\rm phot}\approx3.27$), the lowest-redshift spectroscopic LRD in our sample. Including these objects increases the spectroscopic sample to 131 sources while leaving the outlier fraction unchanged.

\begin{figure}[ht!]
    \centering
    \includegraphics[width=1.0\linewidth]{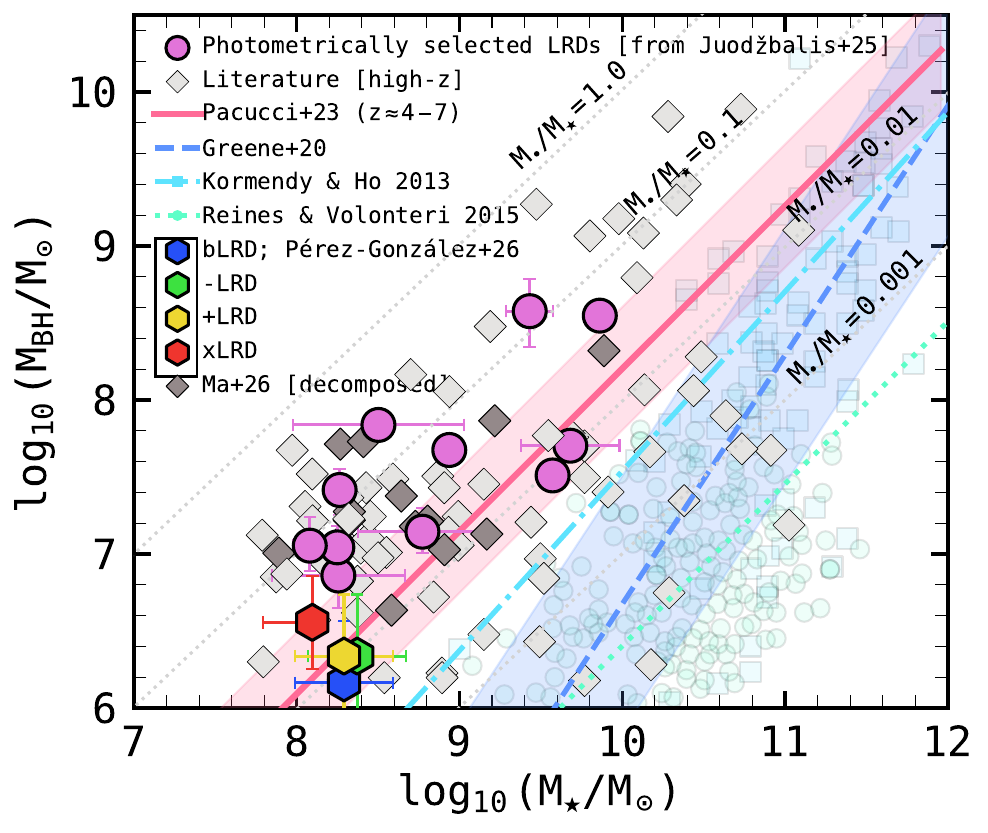}
    \caption{Black hole mass ($M_{\bullet}$) as a function of stellar mass ($M_{\bigstar}$) for the photometrically selected LRDs from \citet{juodzbalis_jades_2025}, compared with measurements from recent studies at similar redshifts (light gray diamonds; \citealt{matsuoka_discovery_2019, furtak_jwst_2023, harikane_jwstnirspec_2023, kocevski_hidden_2023, kokorev_uncover_2023, larson_ceers_2023, maiolino_jades_2023, maiolino_small_2024, matthee_little_2024, rinaldi_not_2025, juodzbalis_jades_2025}) and AGN–host decomposed estimates from \citet{ma_undermassive_2026} (dark gray diamonds). We also show the $M_{\bullet}$ measurements for different LRD sub-types from \citet{perez-gonzalez_little_2026}, together with the local $M_{\bigstar}$–$M_{\bullet}$ relations (\citealt{kormendy_coevolution_2013, reines_relations_2015, greene_intermediate-mass_2020}) and the high-redshift relation proposed by \citet{pacucci_jwst_2023}.}  
    \label{fig:mbh_spec_ignas}
\end{figure}

Finally, we cross-match our sample with the black hole mass compilation of \citet{juodzbalis_jades_2025}, finding 11 matches with secure $M_{\bullet}$ estimates out of their 34 Type~1 AGNs. These span a wide range of redness, from moderately red ($\approx0.5$ mag) to extremely red ($\gtrsim1.5$ mag) under our adopted color criterion, and therefore also exhibit diverse spectral shapes, from curved SEDs to power-law-like continua, consistent with recent results from \citet{barro_cliff_2025}. As shown in Figure~\ref{fig:mbh_spec_ignas}, these photometrically selected LRDs follow the same trend seen in recent JWST studies, lying systematically above the local scaling relations (\citealt{kormendy_coevolution_2013, greene_intermediate-mass_2020}).

\subsubsection{Spectral diversity of photometrically selected LRDs}

Recently, \citet{barro_cliff_2025} assembled one of the largest homogeneous samples of photometrically selected LRDs with JWST/NIRSpec spectroscopy (118 sources), showing that LRDs populate a continuous sequence in rest-frame UV–optical parameter space spanning nearly four magnitudes in optical color and a wide range of spectral shapes. In particular, \citet{barro_cliff_2025} highlighted that the characteristic “V-shaped” SED of LRDs actually spans a continuous sequence between systems with strong Balmer breaks (e.g., the Cliff, RUBIES-BLAGN-1, A2744-QSO1, and MoM-BH*-1; \citealt{de_graaff_remarkable_2025, wang_rubies_2024, furtak_jwst_2023, naidu_black_2025}) and sources with smooth power-law–like continua (e.g., {\it Virgil}; \citealt{iani_midis_2024, rinaldi_deciphering_2025}). This behavior likely indicates that LRDs trace different regimes of AGN and host-galaxy dominance rather than forming a single physical class. 

Consistent with this picture, \citet{perez-gonzalez_little_2026}, using a larger sample of 249 NIRSpec/PRISM spectra, find ordered trends in continuum shape and spectral features. In particular, they find that “bluer” LRDs are characterized by UV continua largely dominated by stellar emission, while progressively redder systems display increasing optical curvature and stronger Balmer breaks, together with a growing prominence of AGN-related signatures such as Fe\,{\sc ii} complexes, broad Balmer emission, and hot-dust emission at rest-frame near-infrared wavelengths (when considering MIRI data).

\begin{figure}[ht!]
    \centering
    \includegraphics[width=\linewidth]{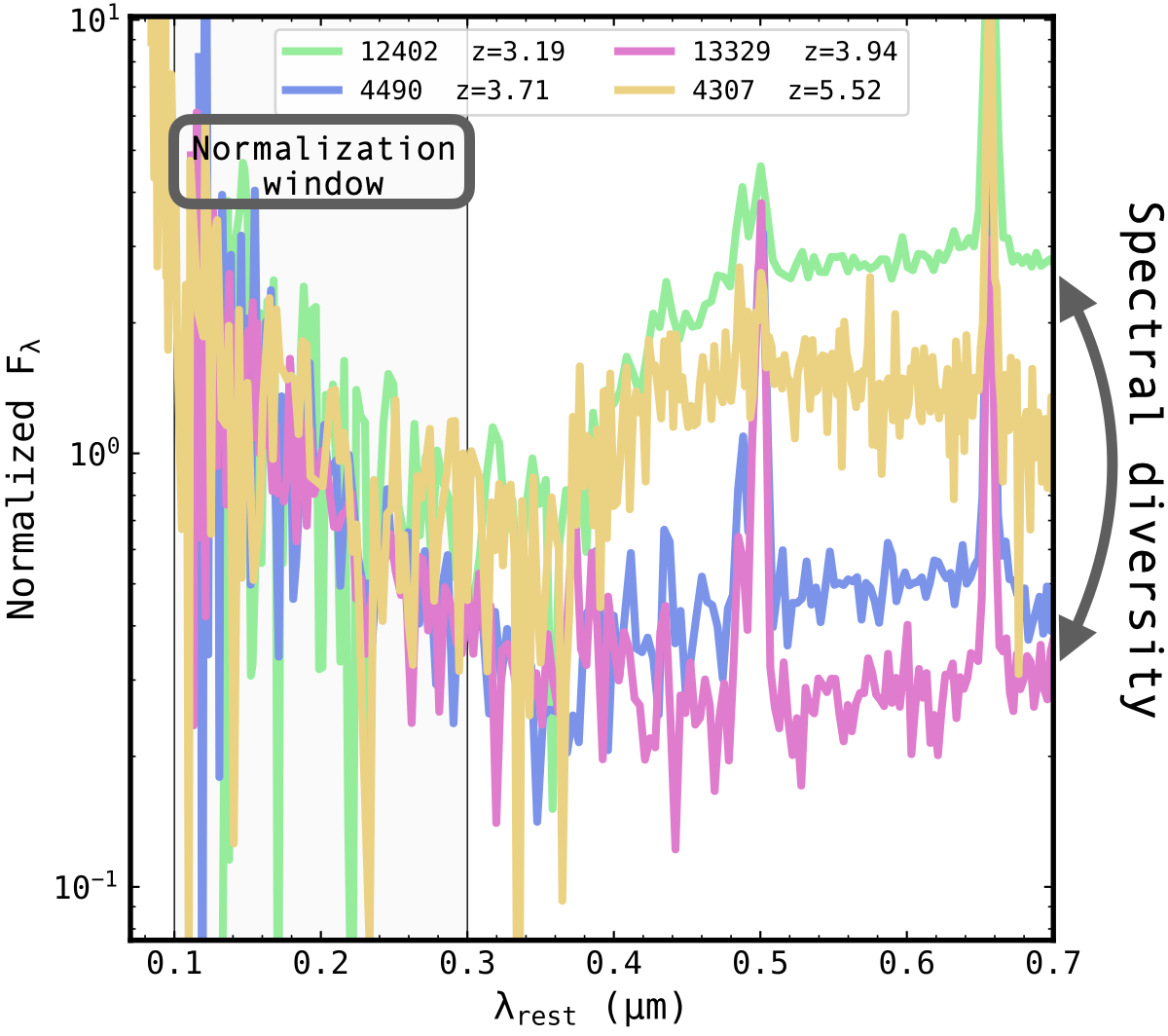}
    \caption{Example of spectral diversity among the photometrically selected LRDs with NIRSpec/PRISM data (four sources are shown for clarity). Spectra are normalized at $\lambda_{\rm rest}=0.1\text{--}0.3\,\rm \mu m$ to highlight how the rest-frame optical varies significantly among photometrically selected LRDs.}
    \label{fig:spectral_diversity}
\end{figure}

Together, these studies show that LRDs form a heterogeneous population spanning a continuous range of properties, and that restrictive photometric selection criteria can bias samples toward one extreme. Indeed, \citet{perez-gonzalez_little_2026} find that the reddest systems (their “extreme” LRDs) represent only $\approx12\%$ of their sample, highlighting the need to fully cover this sequence to constrain the nature of these sources and their evolution along it. An example of this spectral diversity among our photometrically selected LRDs is shown in Figure \ref{fig:spectral_diversity}. In the next section, we will see that the photometrically selected LRDs indeed form a continuous sequence in color–luminosity space, where redder systems exhibit higher optical luminosities, while less red systems are associated with lower optical luminosities.

\subsection{General properties of the photometrically selected LRDs}

Our selection on compactness requires $R_{\rm eff, F444W}\lesssim0.06$\arcsec, ensuring that only unresolved or nearly unresolved sources are included. When examining the resulting physical sizes, the selected LRDs are systematically more compact than the discarded galaxies, with effective radii on average $\approx1.6\times$ smaller. Our LRD sample has a median $R_{\rm eff}\approx125\, \rm pc$ in F444W (and even more compact, following the discussion in \citealt{whalen_limitations_2025}). These sizes are significantly smaller than the typical effective radii of star-forming galaxies at high redshift (e.g., $z\gtrsim5$; \citealt{langeroodi_evolution_2023, ormerod_epochs_2024}) and are consistent with recent size measurements for LRDs reported in the literature (e.g., \citealt{kokorev_census_2024, baggen_small_2024, furtak_jwst_2023}).

Moreover, our full LRD sample spans a broad range in UV luminosity: the bulk of the population (16th–84th percentile) lies between $M_{\rm UV}\approx-19$ and $\approx-17$, while the full population extends to fainter ($M_{\rm UV}\approx-16$) and brighter ($M_{\rm UV}\approx-21$) magnitudes. Here, $M_{\rm UV}$ is obtained by fitting the observed photometry tracing the rest-frame $1300$–$2500\,\text{\AA}$ (assuming $f_{\lambda}\propto \lambda^{\beta}$) continuum and evaluating the best-fit model at $1500\,\text{\AA}$.

To explore their stellar properties, we employed {\sc bagpipes}\footnote{We make use of photometry carried out with {\tt CIRC3} from JADES DR5.}. In particular, we decided to adopt a simple phenomenological AGN component only, following \citet{rinaldi_not_2025} (see their Section~4.1).
Briefly, \textsc{bagpipes} fits were performed using \citet{bruzual_stellar_2003} stellar population models with a \citet{kroupa_variation_2001} IMF (100 $M_{\odot}$ cutoff) and nebular emission modeled with \textsc{cloudy} \citep{ferland_2013_2013}. We adopted a continuity non-parametric SFH \citep{leja_how_2019}, with age bins defined in look-back time from $z=30$ to the age of the Universe at each source redshift. Stellar masses were allowed to vary between $10^5$ and $10^{13}\,M_{\odot}$ (log-uniform prior), $A_V$ between 0–6 assuming a \citet{calzetti_dust_2000} law, and metallicity between $0$–$2.5\,Z_{\odot}$, while the ionization parameter (U) was allowed to vary between -4 and -0.001. As discussed in \citet{rinaldi_not_2025}, fits that include only a stellar component tend to yield systematically lower values of both $A_V$ and $M_{\bigstar}$; to first order, the reduced $M_{\bigstar}$ primarily reflects the lower inferred extinction. 

Using the AGN-inclusive model, we infer $A_V \approx 2.74_{-1.40}^{+1.34}$ (16th–84th percentiles), consistent with recent spectroscopic studies (e.g., \citealt{nikopoulos_evidence_2025}). Moreover, these values are fully consistent with those reported in \citet{labbe_population_2023} and in other recent studies of LRDs based on photometry (e.g., \citealt{kocevski_rise_2025, kokorev_census_2024, akins_cosmos-web_2025, rinaldi_not_2025}).

We also estimated the bolometric luminosity ($L_{\rm Bol}$) of our photometrically selected LRDs. $L_{\rm Bol}$ was derived from the dust--corrected best--fit SED using the monochromatic luminosity at 5100~\AA\; and adopting a bolometric correction factor of 9 (e.g., \citealt{richards_spectral_2006}). Despite the non-ideal nature of this assumption, which attributes the entirety of the observed red continuum flux to AGN emission, we adopt this approach to remain consistent with most recent literature compilations of AGN bolometric luminosities. Nonetheless, it is important to note that \citet{greene_what_2026} recently suggested that commonly adopted estimates of $L_{\rm Bol}$ for LRDs may be overestimated. Under the assumption that the rest-frame optical emission is intrinsically red and not significantly affected by dust attenuation, they find that most of the luminosity emerges in the rest-frame optical and derive $L_{\rm Bol}/L_{5100} \approx 5$, roughly a factor of two lower than the canonical AGN correction. As a consequence, in their framework, $L_{\rm Bol}$ inferred using standard prescriptions can be overestimated by up to an order of magnitude.

\begin{figure}[ht!]
    \centering
    \includegraphics[width=1.0\linewidth]{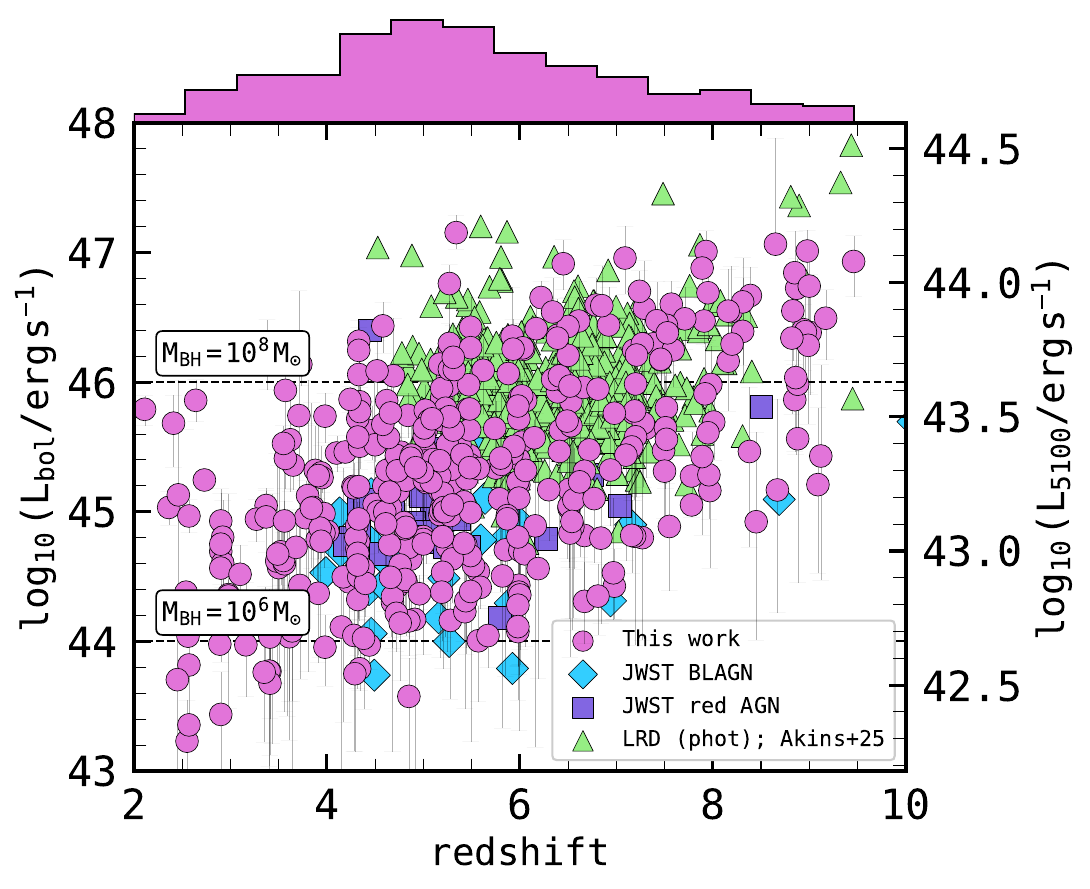}
    \caption{$L_{\rm Bol}$ as a function of redshift. A histogram of the redshifts is shown at the top of the panel.  For context, we include the LRD sample from \citet{akins_cosmos-web_2025} together with other recent studies, grouped into confirmed broad-line AGNs (\citealt{larson_ceers_2023, harikane_jwstnirspec_2023, maiolino_jades_2023, ubler_ga-nifs_2023, bogdan_evidence_2024, maiolino_small_2024, parlanti_ga-nifs_2024, ubler_ga-nifs_2024}) and red AGNs (\citealt{kokorev_uncover_2023, furtak_high_2024, greene_uncover_2024, matthee_little_2024}). Horizontal dashed lines indicate the expected $L_{\rm Bol}$ for black hole masses in the range $\log_{10}(M_{\bullet}/M_{\odot})=6$–8 under the assumption of Eddington-limited accretion. The right y-axis shows the $L_{5100}$ range spanned by our sample, which is consistent with the distribution measured for 249 spectroscopically confirmed LRDs by \citealt{perez-gonzalez_little_2026}.}  
    \label{fig:Lbol_vs_z}
\end{figure}

We find that our sample occupies a similar region of parameter space as the LRD samples presented by \citet{akins_cosmos-web_2025} and \citet{kokorev_census_2024}, with only a small fraction of sources ($\lesssim2\%$) exceeding bolometric luminosities of $L_{\rm Bol}\approx10^{47}\,\mathrm{erg\,s^{-1}}$. Overall, our sample spans $L_{\rm Bol}\approx10^{43\text{--}47}\,\mathrm{erg\,s^{-1}}$, consistent with recent literature derived under similar assumptions. In Figure \ref{fig:Lbol_vs_z}, we show our measurements in the context of recent results about LRDs from other studies under similar assumptions (e.g., \citealt{akins_cosmos-web_2025, kokorev_census_2024}). In contrast, the observed $L_{5100}$ values (see right y-axis of Figure ~\ref{fig:Lbol_vs_z}) span a range consistent with that reported by \citet{perez-gonzalez_little_2026} for a sample of 249 spectroscopically confirmed LRDs, specifically $L_{5100}\approx10^{42\text{--}45}\,\rm erg\,s^{-1}$. 

\begin{figure*}[ht!]
    \centering
    \includegraphics[width=1.0\linewidth]{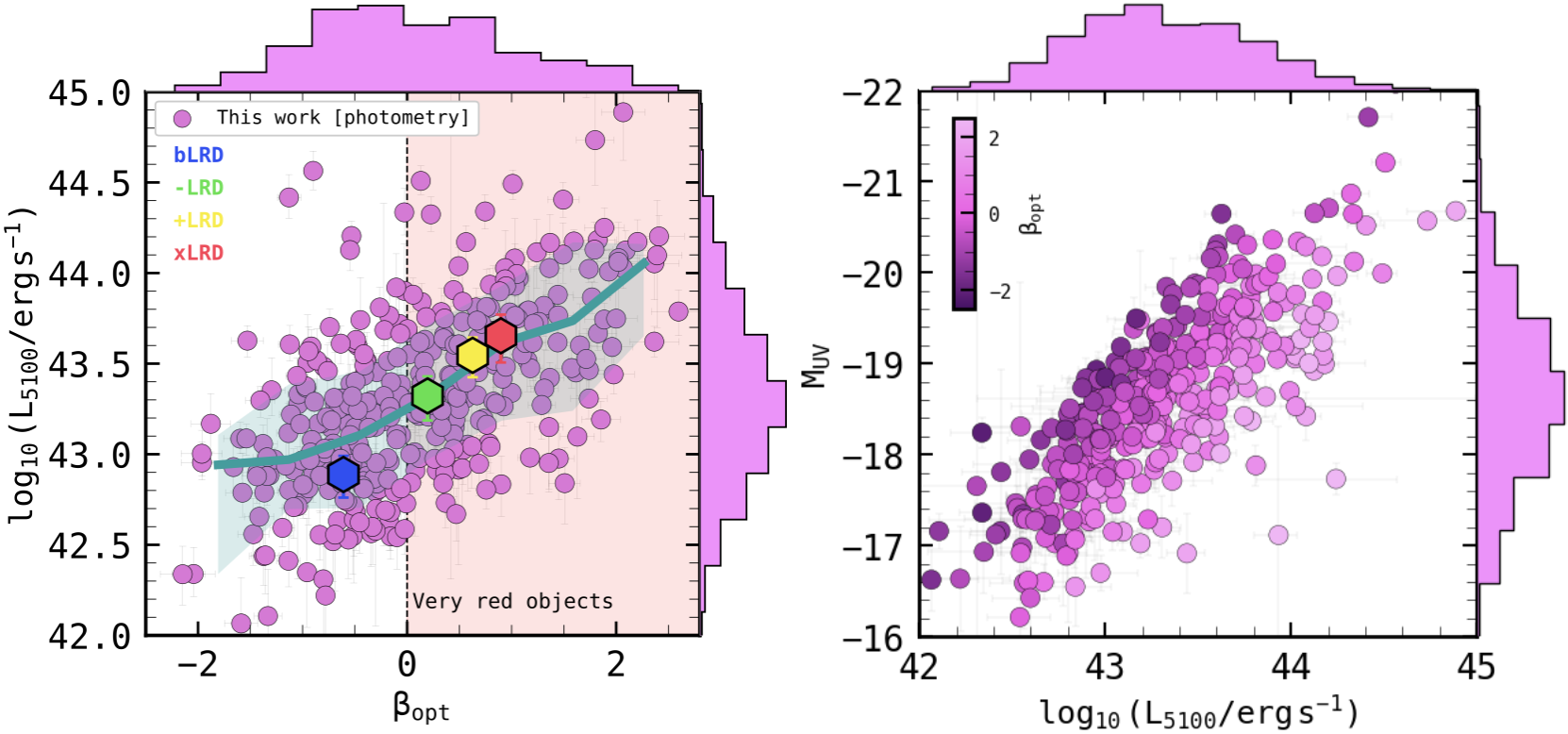}
    \caption{{\bf Left panel:} $\beta_{\rm opt}$ as a function of $L_{5100}$ for the photometrically selected LRDs. As expected, a clear, broad positive correlation is observed (Spearman $\rho=0.59$), with a scatter of $\sigma\approx0.40$ dex around the best-fit relation. For reference, we show the trends inferred from the templates of \citet{perez-gonzalez_little_2026}, which define four LRD sub-types based on stacked NIRSpec/PRISM spectra. The vertical line marks the boundary of commonly adopted selection criteria (e.g., \citealt{kocevski_rise_2025}), separating typically selected LRDs from less red systems. Notably, the bLRD (blue) and part of the $-$LRD (green) population would fall below this threshold, implying that standard selections may miss up to $\approx50\%$ of the identified sub-types. {\bf Right panel:} $M_{\rm UV}$ as a function of $L_{5100}$, color-coded by $\beta_{\rm opt}$. Similarly to the plot on the left, a positive correlation is also present, while at fixed $M_{\rm UV}$ a wide range of $\beta_{\rm opt}$ (and thus $L_{5100}$) is observed, highlighting the heterogeneous nature of LRDs (at fixed $M_{\rm UV}$)  and its potential impact on their inferred demographic evolution.}
    \label{fig:L5100_beta_opt}
\end{figure*}

As expected, we find a broad positive correlation (Spearman coefficient is $\rho=0.59$, and the whole distribution shows, on average, a scatter of $\sigma \approx 0.40$ dex) between $L_{5100}$ and $\beta_{\rm opt}$\footnote{We adopt $f_{\lambda}\propto\lambda^{\beta}$ to derive $\beta_{\rm opt}$, following \citet{kocevski_rise_2025}.} (Figure~\ref{fig:L5100_beta_opt}, left panel), measured over $0.3\text{--}0.8\,\rm \mu m$ in the rest frame, such that redder slopes correspond to higher $L_{5100}$. We highlight that this broad correlation persists when using colors (e.g., $\rm F277W\text{--}F444W$) instead of $\beta_{\rm opt}$, the latter adopted here to demonstrate that the relation is independent of redshift.  This relation traces a continuous sequence across photometrically selected LRDs, from the least to the most red systems, consistent with the trend reported by \citet{perez-gonzalez_little_2026} in $L_{5100}/L_{2500}$ as a function of $L_{5100}$ (their Figure 7). For comparison, we consider the four representative LRD templates of \citet{perez-gonzalez_little_2026}, spanning from blue (bLRD) to extremely red (xLRD), and estimate their $L_{5100}$ and $\beta_{\rm opt}$, finding that they follow the same trend as our photometrically selected LRDs. In particular, restricting the selection to only the very red sources (as commonly done in classic LRD selections) would exclude up to half of the sub-types identified by \citet{perez-gonzalez_little_2026}, as they would not satisfy standard selection thresholds. This would bias our view of the LRD population toward a specific sub-branch, potentially leading to models that capture only a subset of the phenomenon rather than its full diversity. 

As expected, a broad positive correlation is also present between $M_{\rm UV}$ and $L_{5100}$ (Figure~\ref{fig:L5100_beta_opt}, right panel), such that more UV-luminous sources exhibit higher $L_{5100}$. Crucially, at fixed $M_{\rm UV}$, LRDs span the full range of optical redness (from less red to very red), indicating that UV-selected samples encompass a heterogeneous population; this mixing of distinct sub-types can bias the inferred UV luminosity functions and their evolution across cosmic time.

We also estimate an order-of-magnitude black hole mass by assuming Eddington-limited accretion for our LRD candidates, i.e., $L_{\rm Bol}\approx L_{\rm Edd}$, where $L_{\rm Edd}$ scales linearly with black-hole mass. Under this assumption, the black-hole mass is given by $M_{\rm \bullet}=L_{\rm Bol}/(1.26\times10^{38}\,{\rm erg\,s^{-1}})\,M_\odot$, where the normalization follows the canonical expression for the Eddington luminosity (e.g., \citealt{rybicki_radiative_1979, peterson_introduction_1997}). Applying this conversion to our dust-corrected $L_{\rm Bol}$ values yields a median $\log_{10}(M_{\bullet}/M_\odot)\approx7.25^{+0.40}_{-0.36}$ (16th–84th percentiles)\footnote{This should be regarded as a lower limit, since we assume the Eddington limit as the maximum allowed accretion rate.}. We find that such value is consistent with the average $M_{\bullet}$ estimate obtained for the subset of our photometrically selected LRDs after cross-matching our sample with the black hole mass measurements reported by \citet{juodzbalis_jades_2025} (see Figure \ref{fig:mbh_spec_ignas})\footnote{Using instead the scaling relation of \citealt{greene_intermediate-mass_2020} together with the median $M_\bigstar$ implies masses lower by approximately an order of magnitude. We note that $L_{\rm Bol}$ traces the accretion rate and radiative efficiency, so the inferred $M_{\bullet}$ should be regarded as an order-of-magnitude estimate}.  

\begin{figure}[ht!]
    \centering
    \includegraphics[width=1\linewidth]{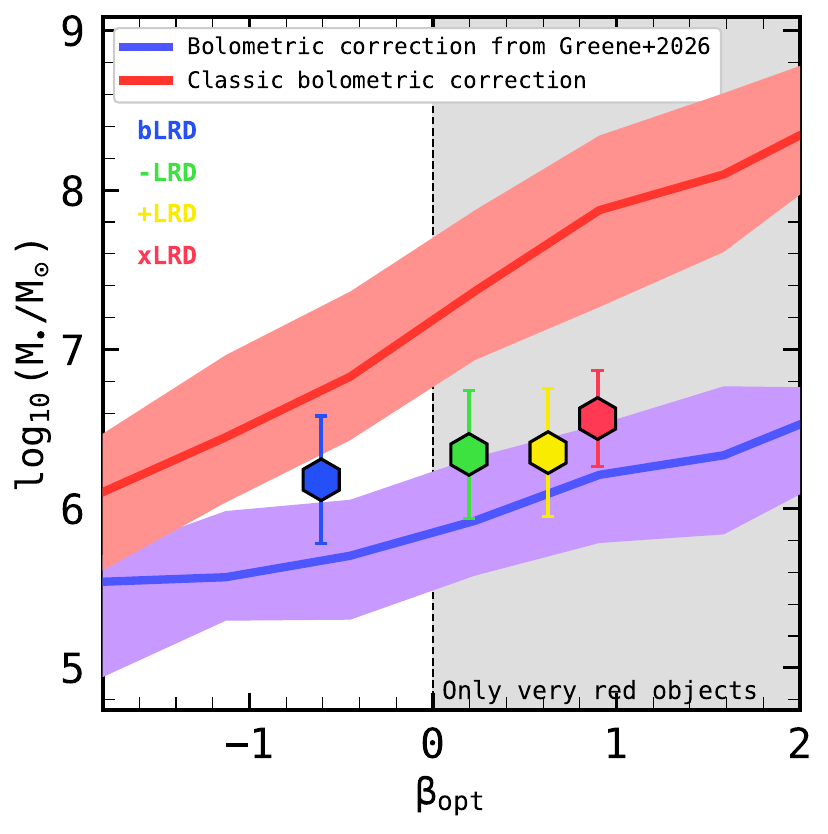}
    \caption{$M_{\bullet}$ as a function of $\beta_{\rm opt}$, comparing estimates derived from dust-corrected $L_{\rm Bol}$ with those obtained under the assumptions of \citealt{greene_what_2026}, in both cases assuming $L_{\rm Bol}\approx L_{\rm Edd}$. A clear positive trend is observed in both approaches, with redder sources corresponding to higher $M_{\bullet}$, although the classical prescription yields systematically larger masses. For reference, we include the $M_{\bullet}$ estimates for the four LRD sub-types identified by \citet{perez-gonzalez_little_2026}, which follow the same trend and are more consistent with the \citealt{greene_what_2026} scenario.}
    \label{fig:MBH_comparison}
\end{figure}

If a significant fraction of the bolometric output arises from star formation, this estimate may be overestimated, as would also be the case if the true bolometric correction is lower than typically adopted in the literature (see discussion in \citealt{greene_what_2026}). For the latter case, Figure \ref{fig:MBH_comparison} compares $M_{\bullet}$ derived from dust-corrected $L_{\rm Bol}$ with estimates obtained under the assumptions of \citealt{greene_what_2026} (in both cases, assuming  $L_{\rm Bol}\approx L_{\rm Edd}$) as a function of $\beta_{\rm opt}$, which shows that redder sources correspond to higher $M_{\bullet}$ in both cases. However, the classical approach yields systematically larger masses than the prescription of \citealt{greene_what_2026}. For reference, we include the $M_{\bullet}$ estimates for the four LRD sub-types identified by \citealt{perez-gonzalez_little_2026}, which follow the same trend and are more consistent with the \citealt{greene_what_2026} scenario (which should represent a lower limit given our assumption).  In any case, our values are consistent with those reported in similar studies, such as \citet{kokorev_census_2024} and \citet{akins_cosmos-web_2025}, under similar assumptions. 

\section{The number density of Little Red Dots in the JADES fields}

The primary motivation of this work is to assess how the inferred LRD population changes under more inclusive photometric selection criteria. The JADES fields provide an optimal dataset for this experiment, given their exceptional JWST depth and extensive multiwavelength coverage, enabling robust constraints on the LRD luminosity function over a relatively large blank-field area. Nevertheless, cosmic variance may still affect number densities, particularly at the highest redshifts; therefore, we regard this as a first step toward a more comprehensive census of LRDs, including other legacy fields.

\subsection{UV Luminosity Function}

\begin{figure*}[ht!]
    \centering
\includegraphics[width=0.98\textwidth]{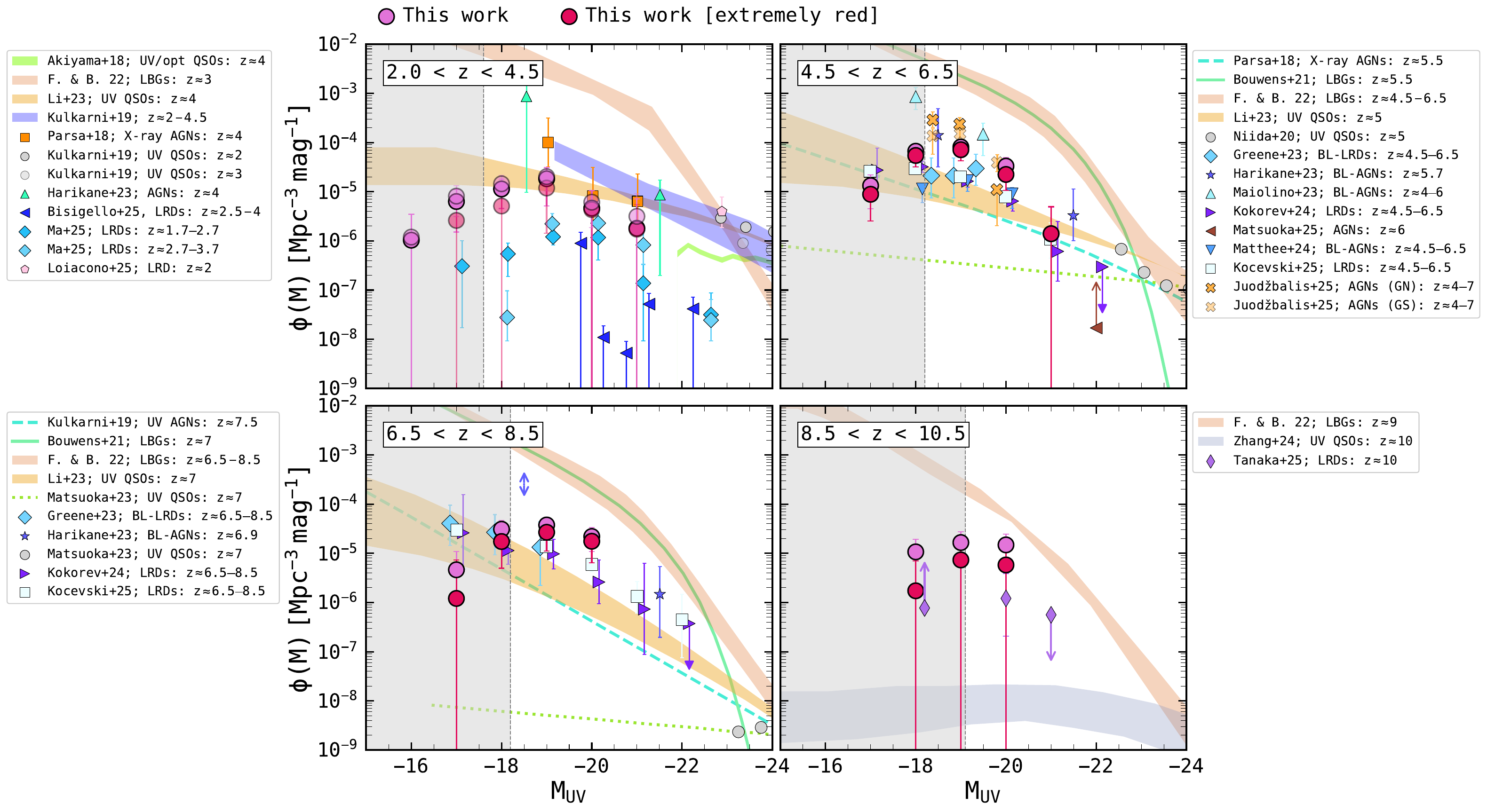}
    \caption{The UV luminosity function of our photometrically selected LRD sample in the JADES fields is presented in four redshift bins and compared with existing observational constraints: $z=2\text{--}4.5$ (\citealt{parsa_no_2018, kulkarni_evolution_2019, harikane_jwstnirspec_2023,bisigello_euclid_euclid_2025, ma_counting_2025, loiacono_big_2025}), $z=4.5\text{--}6.5$ (\citealt{niida_faint_2020, greene_uncover_2024, harikane_jwstnirspec_2023,
    kokorev_census_2024, maiolino_jades_2023, matsuoka_shellqs_2025, matthee_little_2024, kocevski_rise_2025, juodzbalis_jades_2025}), $z=6.5\text{--}8.5$ (\citealt{greene_uncover_2024, harikane_jwstnirspec_2023, matsuoka_quasar_2023, kokorev_census_2024, kocevski_rise_2025}), and $z=8.5\text{--}10.5$ (\citealt{tanaka_discovery_2025}). 
    Filled circles in magenta represent sources selected using our primary selection criteria. Semi-transparent circles show the effect of including additional sources selected with the tailored low-redshift criteria (Sections 3.2 and 3.3). This modification predominantly affects sources at $z\lesssim5$, where the F090W-anchored selection is designed to better cover the UV for LRDs. Red filled circles indicate sources selected using a stricter red color cut (i.e., $>1.5$~mag; \citealt{barro_extremely_2024, akins_cosmos-web_2025}). We note that the primary selection (filled circles), based on the $\rm F277W\text{--}F444W$ red cut, fails to recover sources at $z\lesssim4.5$, whereas the $\rm F200W\text{--}F444W$ criterion (semi-transparent circles) successfully does.   We compare our measurements also with theoretical predictions for both AGNs/QSOs and star-forming galaxies (\citealt{akiyama_quasar_2018, parsa_no_2018, kulkarni_evolution_2019, bouwens_new_2021, finkelstein_coevolution_2022, li_reconstruction_2024, zhang_trinity_2024}). Finally, the light gray shaded region marks where the UV LF becomes incomplete at each redshift.} 
    \label{fig:UVLF}
\end{figure*}

To estimate the UV luminosity function (LF) of the photometrically selected LRDs, we adopt an approach based on the classical \(V_{\rm max}\) estimator \citep{schmidt_space_1968}, generalized to account for the redshift dependence of the photometric selection of LRDs. In its standard formulation, the $V_{\rm max}$ method assigns to each source the maximum comoving volume over which it could be observed given the survey depth. In our case, however, the detectability of a source depends not only on its apparent brightness and spectral shape, but also on the applied color-based selection criteria.

For this reason, instead of defining a single maximum observable redshift, we compute for each source an effective volume $V_{{\rm eff},i}$ by applying the same, fixed selection criteria to the source’s expected photometry at different redshifts. Starting from the observed photometry at the best-fit photometric redshift, we estimate how the source fluxes would appear at higher redshifts by accounting for cosmological dimming and bandpass shifting (the K-correction; e.g., \citealt{oke_energy_1968}), and we re-evaluate whether the source continues to satisfy the adopted detection and color cuts. This defines a source-specific selection function $S_i(z)$, which is unity when the source satisfies the selection and zero otherwise. The effective volume is then given by
\begin{equation}
    V_{{\rm eff},i} = \int_{z_1}^{z_2} S_i(z)\,\frac{{\rm d}V}{{\rm d}z}\,{\rm d}z,
\end{equation}
where $(z_1,z_2)$ denotes the redshift bin under consideration and ${\rm d}V/{\rm d}z$ is the differential comoving volume element for the corresponding survey area. The luminosity function is finally computed as
\begin{equation}
    \Phi(x) = \frac{1}{\Delta x} \sum_i \left[ V_{{\rm eff},i} \right]^{-1},
\end{equation}
where $\Delta x$ is the bin width in the quantity $x$ (i.e., $M_{\rm UV}$). In particular, we verified that our inferred $\Phi(x)$ is consistent with the result obtained using the classical $1/V_{\rm max}$ estimator used in recent works (e.g., \citealt{kokorev_census_2024}), finding no statistically significant differences between the two approaches. Moreover, following \citet{kokorev_census_2024}, we do not apply a global magnitude-completeness correction, as this would require assumptions on the intrinsic source population. Instead, we estimate the median $M_{\rm UV}$ $5\sigma$ completeness limits based on the depths of the filters probing rest-frame $\approx1500$ \AA\; at each redshift, requiring a detection at $\mathrm{S/N}>5$.

Uncertainties on $\Phi(x)$ are estimated by combining Poisson counting errors, computed following \citet{gehrels_confidence_1986}, with the contribution from photometric-redshift uncertainties. The latter is quantified through Monte Carlo sampling of each source’s $p(z)$ distribution \citep{marchesini_evolution_2009}. Given the relatively limited survey area, we further include a conservative cosmic-variance contribution to the total error budget based on \citet{trenti_cosmic_2008}.

In Figure~\ref{fig:UVLF}, we present our measurements of the UV luminosity function of LRDs in four redshift bins: $z\approx2\text{--}4.5$, $z\approx4.5\text{--}6.5$, $z\approx6.5\text{--}8.5$, and $z\approx8.5\text{--}10.5$. The absolute UV magnitude, $M_{\rm UV}$, is measured at 1500\,\AA, consistent with the convention adopted for blue quasars. The redshift binning is chosen to both facilitate direct comparison with existing studies and to minimize the impact of photometric redshift uncertainties on the derived luminosity functions. In each panel, we present the UV luminosity function of photometrically selected LRDs using our selection criteria, and compare it with the UV LF obtained when applying stricter criteria for $\rm F277W\text{--}F444W$, which require each source to be redder than 1.5 mag in this case (see \citealt{barro_extremely_2024, akins_cosmos-web_2025}). We show our results as filled circles for the primary selection criteria, while semi-transparent circles are shown only in the first panel of Figure~\ref{fig:UVLF} to illustrate the impact of the alternative selection strategy, which preferentially selects sources at $z<5$ and therefore has a negligible effect on the higher-redshift bins (see Section 3.2 and 3.3 for the selection criteria adopted for our sample).  

As already noted in previous studies (e.g., \citealt{kokorev_census_2024}), comparisons across the literature must account for the intrinsic difficulty of defining an accurate selection function for spectroscopically confirmed samples, and hence of deriving robust $V_{\max}$ corrections. We also remind the reader that our sample is selected purely via photometry, and that the $V_{\max}$ correction is computed solely from the adopted selection criteria, following the procedure outlined in \citet{kokorev_census_2024}. This conservative approach is designed to avoid significantly overestimating the number counts and, in turn, misrepresenting the true abundance of the selected LRDs.

At $z\approx2\text{--}4.5$ (Figure~\ref{fig:UVLF}; top left panel), the only available LRD comparisons are \citet{ma_counting_2025} and \citet{bisigello_euclid_euclid_2025} (plus a single source from \citealt{loiacono_big_2025}). Both adopt different selection criteria and probe much larger areas, but rely on data significantly shallower than that used in this work. \citet{ma_counting_2025} report a decline in LRD number density toward lower redshift, from $(4.6\pm1.8)\times10^{-6}\,\mathrm{cMpc}^{-3}$ at $z\approx2.7\text{--}3.7$ to $(2.1\pm1.1)\times10^{-6}\,\mathrm{cMpc}^{-3}$ at $z\approx1.7\text{--}2.7$, corresponding to nearly an order-of-magnitude drop from $z>4$ to $z<4$. Our measurements (primary selection; magenta) lie systematically above both \citet{ma_counting_2025} and \citet{bisigello_euclid_euclid_2025}, likely due to our broader redshift bin, with $\approx43\%$ of sources at $z\approx3.7\text{--}4.5$, outside the range probed by \citet{ma_counting_2025}. Overall, we find broad agreement with theoretical expectations at similar redshifts. Including sources selected with the tailored low-$z$ LRD criteria (Section~3.3; semi-transparent circles in Figure~\ref{fig:UVLF}) yields results consistent, within uncertainties, with the primary selection. Their inclusion does not affect the UV LF at higher redshifts, and they are therefore omitted from the higher-$z$ panels.

Remarkably, adopting a stricter $\mathrm{F277W\text{--}F444W}>1.5$ mag cut within the primary selection (F150W, F200W, F277W, F444W) yields no sources. In contrast, the alternative selection based on F090W, F150W (or F115W), F200W, and F444W recovers photometrically selected LRDs (semi-transparent red). Our measurements remain consistent, within uncertainties, with \citet{ma_counting_2025} at $z\approx2.7\text{--}3.7$. This demonstrates that the inferred results depend strongly on the adopted selection, and that identifying LRDs at later times is highly sensitive to the specific filter combinations probing the rest-frame UV and optical/NIR emission.

At $z \approx 4.5\text{--}6.5$ and $z \approx 6.5\text{--}8.5$ (Figure~\ref{fig:UVLF}; top right and bottom left panels), our measurements are consistent, within uncertainties, with recent LRD studies and, more broadly, JWST-selected AGN samples ( \citealt{harikane_jwstnirspec_2023, maiolino_jades_2023, greene_uncover_2024, kocevski_rise_2025, kokorev_census_2024, matsuoka_shellqs_2025, matthee_little_2024, juodzbalis_jades_2025}), regardless of the adopted $\rm F277W\text{--}F444W$ color cut. These redshift intervals are also the most extensively explored in the current LRD literature, enabling a robust comparison.  Our inferred number densities are comparable to those of X-ray–selected quasars, while extending to higher densities at fainter UV magnitudes, as reported in previous works (e.g., \citealt{kocevski_rise_2025, kokorev_census_2024, akins_cosmos-web_2025}), highlighting a key distinction between LRDs and the pre-JWST quasar population. At these redshifts, our results are also in broad agreement with the UV QSO LF predictions of \citet{li_reconstruction_2024}.

Finally, when moving to $z\approx8.5\text{--}10.5$, we find agreement within the error bars with recent measurements at similar redshifts from \citet{tanaka_discovery_2025} in the COSMOS-Web field. For comparison, we also consider the UV QSO LF at comparable redshifts from \citet{zhang_trinity_2024}, based on simulations, and adopt the conversion from \citet{runnoe_updating_2012} to convert $L_{\rm Bol}$ to $M_{\rm UV}$. We further contrast our results with the predicted galaxy UV LFs from \citet{finkelstein_coevolution_2022}. In this context, our measurements extend those of \citet{tanaka_discovery_2025}, providing the first constraints on the UV LF of LRDs at these redshifts and lying $\approx2$~dex above the expected UV QSO LF at comparable UV luminosities. However, we caution the reader that these results are based solely on photometric redshifts; nonetheless, they provide a useful benchmark for future follow-up studies. We report our measurements in Table \ref{tab:uvlf}.

\begin{deluxetable}{ccc|ccc}
\tablecaption{UV luminosity function of the LRD sample.\label{tab:uvlf}}
\renewcommand{\arraystretch}{1}
\tabletypesize{\small}
\setlength{\tabcolsep}{1pt}
\tablehead{
\multicolumn{3}{c|}{$2.0 < z < 4.5$} & \multicolumn{3}{c}{$4.5 < z < 6.5$} \\
\tableline
\colhead{$M_{\rm UV}$} & \colhead{$\log_{10}(\Phi/(\rm Mpc^{-3}\,mag^{-1}))$} & \colhead{} &
\colhead{$M_{\rm UV}$} & \colhead{$\log_{10}(\Phi/(\rm Mpc^{-3}\,mag^{-1}))$} & \colhead{}
}
\startdata
$-21$ & $-5.74^{+0.60}_{\rightarrow 0}$ & &
$-21$ & $-5.85^{+0.53}_{\rightarrow 0}$ & \\
$-20$ & $-5.33^{+0.36}_{\rightarrow 0}$ & &
$-20$ & $-4.48^{+0.18}_{-0.28}$ & \\
$-19$ & $-4.74^{+0.24}_{-0.55}$ & &
$-19$ & $-4.09^{+0.15}_{-0.20}$ & \\
$-18$ & $-4.94^{+0.22}_{-0.38}$ & &
$-18$ & $-4.18^{+0.14}_{-0.23}$ & \\
$-17$ & $-5.20^{+0.25}_{-0.68}$ & &
$-17$ & $-4.88^{+0.25}_{-0.43}$ & \\
$-16$ & $-5.98^{+0.52}_{\rightarrow 0}$ & &
 & & \\
\tableline
\multicolumn{6}{c}{} \\[-0.2pt]
\tableline
\multicolumn{3}{c|}{$6.5 < z < 8.5$} & \multicolumn{3}{c}{$8.5 < z < 10.5$} \\
\tableline
\colhead{$M_{\rm UV}$} & \colhead{$\log_{10}(\Phi/(\rm Mpc^{-3}\,mag^{-1}))$} & \colhead{} &
\colhead{$M_{\rm UV}$} & \colhead{$\log_{10}(\Phi/(\rm Mpc^{-3}\,mag^{-1}))$} & \colhead{} \\
\tableline
$-20$ & $-4.66^{+0.16}_{-0.33}$ & &
$-20$ & $-4.83^{+0.22}_{-0.46}$ & \\
$-19$ & $-4.42^{+0.16}_{-0.27}$ & &
$-19$ & $-4.78^{+0.18}_{-0.52}$ & \\
$-18$ & $-4.51^{+0.15}_{-0.34}$ & &
$-18$ & $-4.97^{+0.29}_{-0.80}$ & \\
$-17$ & $-5.33^{+0.36}_{\rightarrow 0}$ & &
 & & \\
\enddata
\tablecomments{Measurements refer to the primary selection criteria (Section 3.2 and 3.3).}
\end{deluxetable}

\subsection{Optical Luminosity Function}

Albeit powerful, the UV luminosity function carries significant uncertainties in tracing the true underlying nature of the LRD population across cosmic time. Indeed, despite the growing number of spectroscopic confirmations for these objects (see \citealt{barro_cliff_2025} and \citealt{perez-gonzalez_little_2026} for an overview), the origin of the rest-frame UV emission remains uncertain. It may arise from AGN light (either scattered or transmitted through patchy medium) or from unobscured star formation in the host galaxy. Moreover, in Figure \ref{fig:L5100_beta_opt}, we show that at fixed $M_{\rm UV}$, $\beta_{\rm opt}$ spans a continuous range of values, implying that UV-selected samples encompass the full diversity of LRDs (from less red to extremely red systems). As a result, the UV luminosity function mixes distinct sub-types (\citealt{perez-gonzalez_little_2026}) and physical regimes, complicating its physical interpretation.

For this reason, recent studies have attempted to characterize LRD evolution from a $L_{\rm Bol}$ perspective (see \citealt{kokorev_census_2024} and \citealt{akins_cosmos-web_2025}), which in principle provides a more direct probe of black hole growth.  Nonetheless, several works have highlighted the difficulty of reliably constraining bolometric luminosities for LRDs. When modeled, these systems typically yield large $A_V$ (e.g., \citealt{kokorev_census_2024, akins_cosmos-web_2025, rinaldi_not_2025}), which in turn can bias dust-corrected $L_{\rm Bol}$ estimates toward artificially high values, especially when combined with standard bolometric correction factors.

More recently, an alternative scenario has been proposed in which the observed optical emission ($L_{5100}$) in LRDs is largely intrinsic, rather than dust-reprocessed (see \citealt{greene_what_2026}). Under this assumption, the inferred $L_{\rm Bol}$ would be significantly lower than estimates previous estimates, which in turn can alleviate the tension on the inferred black hole mass estimate under the classic assumption of $L_{\rm Bol}\approx L_{\rm Edd}$ (see Figure \ref{fig:MBH_comparison}).

\begin{figure*}[ht!]
    \centering
\includegraphics[width=0.98\textwidth]{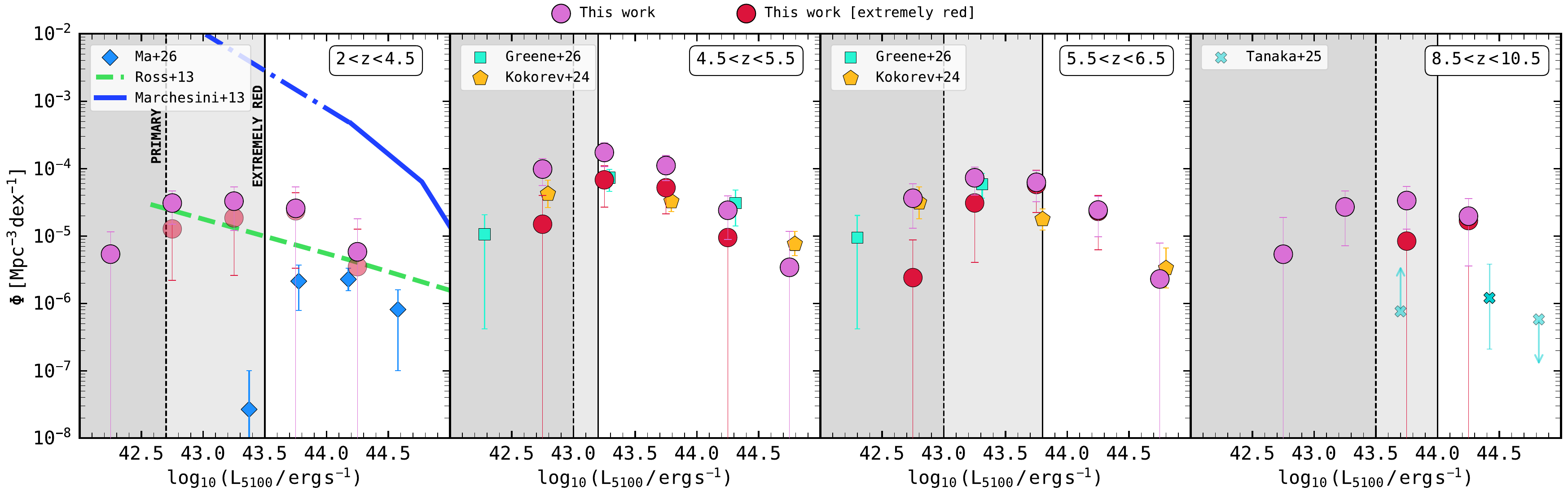}
    \caption{The optical (5100;\AA) luminosity function of LRDs across cosmic time. Magenta circles denote the primary selection, while red circles show the extreme red cut ($\rm F277W\text{--}F444W>1.5$~mag); see section 3.2 and 3.3. The latter misses sources in the lowest redshift bin, whereas the F090W-anchored criterion recovers them (semi-transparent red circles). Given the limited number of optical LF measurements available in the literature, comparisons are restricted to a few studies in each redshift bin. In the lowest redshift bin, we find broad agreement at the bright end with \citet{ma_counting_2025}, where their sample is likely complete. At these redshifts, our LRD sample agrees with the quasar optical LF at $z\approx2.2\text{--}3.5$ (\citealt{ross_sdss-iii_2013}), while lies below the extrapolated optical LF of star-forming galaxies at similar redshift from \citet{marchesini_evolution_2012}. At intermediate redshifts ($z\approx4.5$–$8.5$), our results are consistent with \citet{greene_what_2026} and with an approximate de-correction of $L_{\rm Bol}$ from \citet{kokorev_census_2024}. At the highest redshifts ($z>8.5$), we find overall agreement with the recent compilation of LRD candidates at $z\approx10$ from \citet{tanaka_discovery_2025}, where they are limited only to the bright end of the optical LF.} 
    \label{fig:optical_LF}
\end{figure*}

Given these uncertainties, we do not attempt to study the bolometric LF for our LRD sample. Instead, we report the rest-frame optical luminosity at 5100\,\AA\; (Figure \ref{fig:optical_LF}). We note that constraints on the optical luminosity function for LRDs remain scarce (and more in general for AGNs and SFGs at high redshifts). We therefore compare our measurements with the limited available results in the literature, in particular \citet{ma_counting_2025}, \citet{tanaka_discovery_2025}, and \citet{greene_what_2026}, and perform an approximate de-correction of the estimates from \citet{kokorev_census_2024} to provide an order-of-magnitude comparison. In all cases, we present results for both our primary selection and the complementary extremely red selection (see details in Sections 3.2 and 3.3). Since the extremely red selection yields no detections at $z\lesssim4.5$, we show, in the lowest redshift bin (Figure \ref{fig:optical_LF}; first panel), semi-transparent points obtained using also the refined, F090W-based UV selection (Section 3.3).  At $z>4.5$, differences between the selections are negligible and are therefore not shown. In each panel, we also indicate an empirical $5\sigma$ completeness limit, estimated as the median depth after applying the full set of selection criteria required to classify sources as LRDs. We stress that this estimate should be regarded as conservative, given the difficulty of accurately quantifying completeness for this population in the presence of an uncertain underlying parent distribution.

In the lowest redshift bin, our measurements exceed those reported by \citet{ma_counting_2025} at similar redshifts. This discrepancy likely reflects differences in depth as well as selection criteria. At $z\lesssim4.5$, \citet{ma_counting_2025} rely on either ground-based $K$-band data or Spitzer/IRAC Channel 2 (3.6\ $\mu$m), with $5\sigma$ depths of $\approx23.7$ and $\approx23.5$~mag, respectively. At these redshifts, these bands probe the rest-frame optical emission of LRDs; however, given their depth, they likely sample only the bright end of the population, corresponding to luminosities of $\gtrsim 10^{44}\,\mathrm{erg\,s^{-1}}$, where their sample is expected to be complete. Consequently, despite the large survey area, \citet{ma_counting_2025} are sensitive primarily to the brightest tail of the LRD population; therefore, any comparison must be done with care. In particular, our LRD sample is consistent with the quasar LF at $z\approx2.2\text{--}3.5$ (\citealt{ross_sdss-iii_2013}), and lies below the extrapolated optical LF of star-forming galaxies at similar redshift from \citet{marchesini_evolution_2012}.

In the two intermediate redshift bins ($z\approx4.5\text{--}8.5$), the most relevant comparison is provided by the recent results of \citet{greene_what_2026}, who assume that the optical emission at 5100\,\AA\, is intrinsic. We convert their bolometric luminosities using their revised (lower) bolometric correction and find very good agreement with our measurements. Finally, for illustrative purposes, we approximate a de-correction of the bolometric luminosities reported by \citet{kokorev_census_2024} using their average $A_{V}$, enabling an order-of-magnitude comparison with both our results and \citet{greene_what_2026}; we find overall good agreement.

At the highest redshift bin, the only available comparison is with \citet{tanaka_discovery_2025}. However, this comparison must be treated with caution, as their analysis focuses on a much narrower redshift range around $z\approx10$, whereas our bin spans a broader interval. In addition, their detections rely on relatively shallow MIRI data, which limits their sensitivity to the brightest sources and therefore probes only the bright end of the optical LF at these redshifts. Nevertheless, the results remain broadly consistent within the uncertainties at similar $L_{5100}$.

Overall, because LRDs span a wide range of optical redness at fixed $M_{\rm UV}$, the UV luminosity function mixes intrinsically diverse systems and may not faithfully trace the underlying population, whereas the optical LF provides a more physically consistent view of their evolution across cosmic time. We provide our measurements in Table \ref{tab:optlf}.

\begin{deluxetable}{ccc|ccc}
\tablecaption{Optical (5100\,\AA) luminosity function of the LRD sample.\label{tab:optlf}}
\renewcommand{\arraystretch}{1}
\tabletypesize{\small}
\setlength{\tabcolsep}{1pt}
\tablehead{
\multicolumn{3}{c|}{$2.0 < z < 4.5$} & \multicolumn{3}{c}{$4.5 < z < 6.5$} \\
\tableline
\colhead{$\log_{10}(L_{5100})$} & \colhead{$\log_{10}(\Phi)$} & \colhead{} &
\colhead{$\log_{10}(L_{5100})$} & \colhead{$\log_{10}(\Phi)$} & \colhead{}
}
\startdata
$42.25$ & $-5.27^{+0.33}_{\rightarrow 0}$ & &
 & & \\
$42.75$ & $-4.51^{+0.18}_{-0.31}$ & &
$42.75$ & $-4.01^{+0.16}_{-0.25}$ & \\
$43.25$ & $-4.48^{+0.21}_{-0.44}$ & &
$43.25$ & $-3.76^{+0.14}_{-0.21}$ & \\
$43.75$ & $-4.59^{+0.32}_{\rightarrow 0}$ & &
$43.75$ & $-3.96^{+0.15}_{-0.22}$ & \\
$44.25$ & $-5.23^{+0.49}_{\rightarrow 0}$ & &
$44.25$ & $-4.62^{+0.21}_{-0.43}$ & \\
 & & &
$44.75$ & $-5.47^{+0.53}_{\rightarrow 0}$ & \\
\tableline
\multicolumn{6}{c}{} \\[-0.2pt]
\tableline
\multicolumn{3}{c|}{$6.5 < z < 8.5$} & \multicolumn{3}{c}{$8.5 < z < 10.5$} \\
\tableline
\colhead{$\log_{10}(L_{5100})$} & \colhead{$\log_{10}(\Phi)$} & \colhead{} &
\colhead{$\log_{10}(L_{5100})$} & \colhead{$\log_{10}(\Phi)$} & \colhead{} \\
\tableline
$42.75$ & $-4.44^{+0.22}_{-0.45}$ & &
$42.75$ & $-5.27^{+0.54}_{\rightarrow 0}$ & \\
$43.25$ & $-4.14^{+0.16}_{-0.26}$ & &
$43.25$ & $-4.57^{+0.24}_{-0.58}$ & \\
$43.75$ & $-4.20^{+0.17}_{-0.29}$ & &
$43.75$ & $-4.47^{+0.21}_{-0.42}$ & \\
$44.25$ & $-4.62^{+0.20}_{-0.40}$ & &
$44.25$ & $-4.71^{+0.26}_{-0.74}$ & \\
$44.75$ & $-5.64^{+0.53}_{\rightarrow 0}$ & &
 & & \\
\enddata
\tablecomments{Measurements refer to the primary selection criteria. $L_{5100}$ is unit of $\rm erg\, s^{-1}$ and $\Phi$ is unit of $\rm Mpc^{-3}\, dex^{-1}$.}
\end{deluxetable}

\subsection{Evolution of Little Red Dot number density with cosmic time}

Since our LRD selection spans from Cosmic Noon to the early Universe, we estimate their number density as a function of cosmic time. To explore the evolution of these sources as a function of redshift, we integrate the UV luminosity function down to $M_{\rm UV} = -18.5$, in order to enable a fair and direct comparison with the majority of recent (observational and theoretical) studies in the literature. We find number densities of $(2.82\pm 1.61)\times10^{-5}\,\mathrm{cMpc^{-3}}$ at $z\approx2\text{--}4.5$, $(1.16\pm 0.41)\times10^{-4}\,\mathrm{cMpc^{-3}}$ at $z\approx4.5\text{--}6.5$, $(5.99 \pm2.19)\times10^{-5}\,\mathrm{cMpc^{-3}}$ at $z\approx6.5\text{--}8.5$, and $(3.18\pm 1.28)\times10^{-5}\,\mathrm{cMpc^{-3}}$ at $z\approx8.5\text{--}10.5$. 

\begin{figure*}[ht!]
    \centering
    \includegraphics[width=1.0\linewidth]{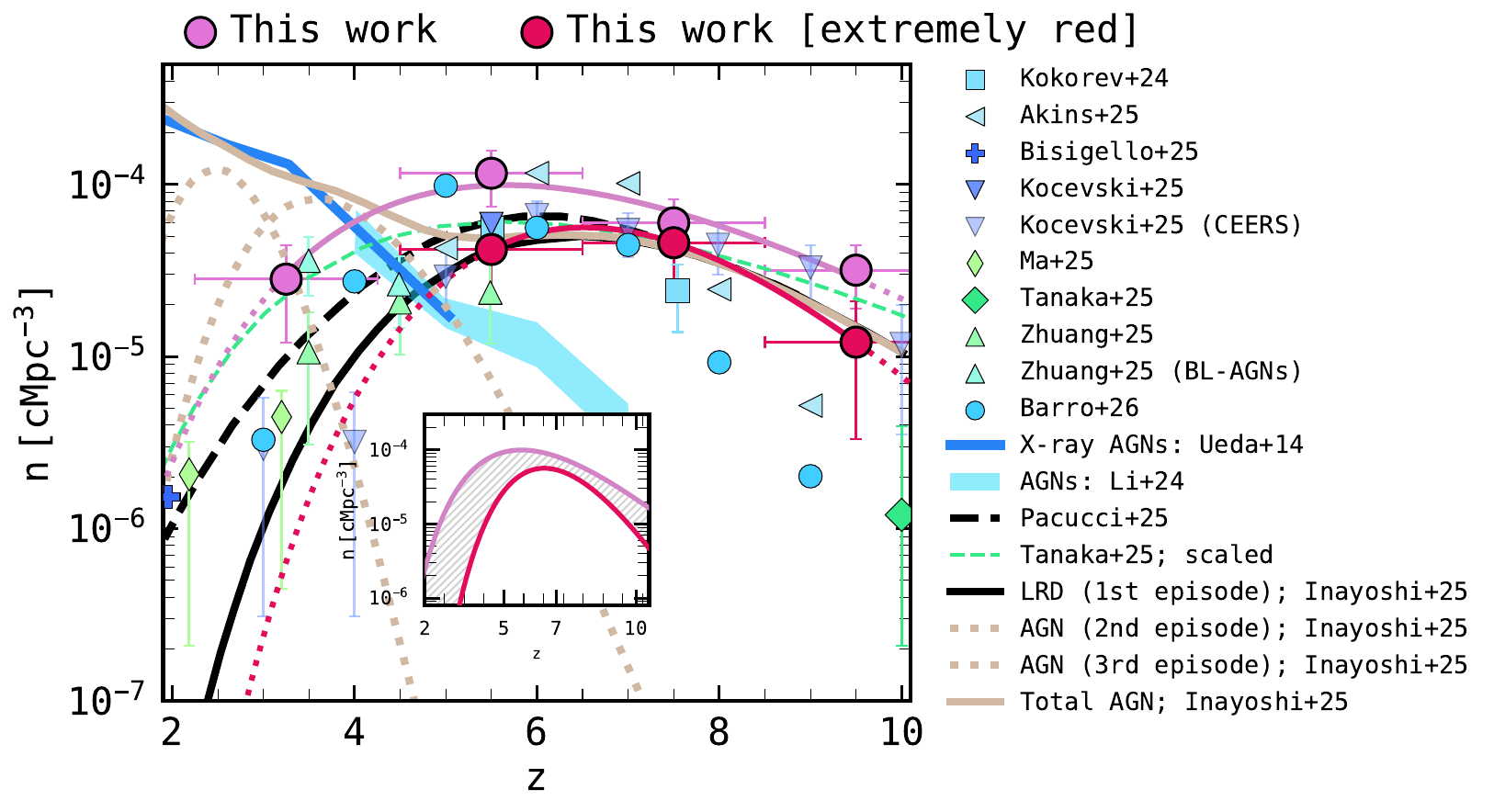}
    \caption{Redshift evolution of the number density of photometrically selected LRDs in the JADES fields (magenta), integrated down to $M_{\rm UV}=-18.5$ to enable a direct comparison with the literature. We include recent observational constraints across cosmic time (\citealt{kokorev_census_2024, akins_cosmos-web_2025, bisigello_euclid_euclid_2025, kocevski_rise_2025, ma_counting_2025, tanaka_discovery_2025, Zhuang_nexus_2025, barro_comprehensive_2026}), together with theoretical predictions from \citet{inayoshi_little_2025} and \citet{pacucci_cosmic_2025}. We also show the AGN evolutionary framework of \citet{inayoshi_little_2025}, in which discrete accretion episodes onto the central black hole are invoked to explain the observed X-ray AGN demographics at $z\lesssim5$ (\citealt{ueda_toward_2014}), with the first episode associated with the LRD phase. Additional predictions include the AGN evolution from \citet{li_reconstruction_2024} and the LRD evolution proposed by \citet{tanaka_discovery_2025}, normalized to \citet{pacucci_cosmic_2025}. We further show our inferred LRD evolution derived from our sample when applying the extreme color cut (red). 
}  
    \label{fig:LRD_evo_z}
\end{figure*}

In Figure~\ref{fig:LRD_evo_z}, we place our results in the context of recent observational determinations. In particular, we also show, in red, the expected number density as a function of cosmic time obtained when applying stricter color cuts commonly adopted in the recent literature, namely $\mathrm{F277W\text{--}F444W}>1.5$ mag \citep{akins_cosmos-web_2025} (see Section 3.2 and 3.3). For reference, we also include the expected redshift evolution of LRDs from recent theoretical predictions by \citet{inayoshi_little_2025} and \citet{pacucci_cosmic_2025}, together with the inferred evolution reported by \citet{tanaka_discovery_2025}, scaled to match \citet{pacucci_cosmic_2025}.

We first focus on the redshift interval most commonly explored in the literature, $z\approx4.5\text{--}8.5$. In this regime, the measurements obtained from our primary selection criteria (magenta) are in broad agreement with recent observational studies (e.g., \citealt{kokorev_census_2024, kocevski_rise_2025}), while remaining systematically higher than the theoretical predictions of \citet{pacucci_cosmic_2025} and \citet{inayoshi_little_2025}.

At lower redshifts, our measurements lie above those reported by \citet{ma_counting_2025}, who relied primarily on ground-based data, but are in agreement with the more recent results obtained by \citet{Zhuang_nexus_2025} for LRDs in the NEXUS survey (\citealt{Zhuang_nexus_2024}). We note that our lowest-redshift bin spans a relatively wide interval, from $z\approx2$ to $z\approx4.5$, with a substantial fraction of the sample located at $z\gtrsim3.7$. As a consequence, comparisons with \citet{ma_counting_2025}, whose measurements are confined to lower redshifts ($z\approx1.7\text{--}3.7$), must be treated with caution. In the highest redshifts probed in this work, $z\approx8.5\text{--}10.5$, our inferred number densities are in broad agreement with the most recent measurements available at these epochs. We note, however, that our measurements lie significantly above those reported by \citet{tanaka_discovery_2025}, which are i) integrated down to $M_{\rm UV}=-20$ and ii) are confined mostly to $z\approx10$.

Interestingly, when adopting stricter color cuts within our primary selection (shown in red), the agreement with the current literature improves significantly, underscoring how the adopted selection strategy shapes the inferred properties of this population. At the same time, these stricter criteria preclude any reliable estimate of the number density at $z\approx2\text{--}4.5$, highlighting the limitations imposed by both the filter choice and the color selection itself. In particular, the filter combination used in the primary selection (optimized for $z\gtrsim4$; i.e., F150W, F200W, F277W, and F444W) becomes progressively less effective at later cosmic times. More generally, restricting the selection to only the reddest sources severely limits our ability to capture the full demographics of the LRD population, biasing the census toward the most extreme objects.

Following \citet{tanaka_discovery_2025}, we also fit the evolution of LRDs based only on the results inferred from our sample integrated down to $M_{\rm UV}=-18.5$. To do so, we follow the log-normal distribution recently proposed by \citet{inayoshi_little_2025}:
\begin{equation}
\begin{aligned}
\Phi_{\mathrm{LRD}}(z) &= \Phi_{0,\mathrm{LRD}}\, f(z)\,
\exp\!\left[
-\frac{\left\{\ln(1+z) - \ln(1+z_0)\right\}^2}{2\sigma_z^2}
\right], \\
f(z) &= \frac{(1+z)^{3/2}}{\left[s(1+z)^{1/2} - 1\right]^2}.
\end{aligned}
\end{equation}
In this framework, $\Phi_{0,\mathrm{LRD}}$ is the normalization of the log-normal distribution, $z_0$ represents the peak redshift, and $\sigma_z$ sets the width of the distribution. The function $f(z)$ accounts for the redshift dependence of both the differential comoving volume element, $\mathrm{d}V/\mathrm{d}z$, and the cosmic time interval, $\mathrm{d}t/\mathrm{d}z$. We adopt $s=0.901$, corresponding to a flat cosmology with $\Omega_{\rm m}=0.3$ and $\Omega_\Lambda=0.7$. For our primary selection criteria (magenta circles), we obtain the following best-fitting parameters are $\Phi_{0,\mathrm{LRD}} = (1.07\pm0.33)\times10^{-5}\,\mathrm{cMpc^{-3}}$, $z_0 = 5.81 \pm 0.39$, and $\sigma_z = 0.28 \pm 0.06$. We find broad agreement with \citet{tanaka_discovery_2025} for the redshift evolution parameters, $z_0$ and $\sigma_z$. However, we find a higher value in $\Phi_{0,\mathrm{LRD}}$, which likely arises from (i) the different selection criteria adopted in this work, with our approach being more inclusive, and (ii) the different integration limits, as we integrate the luminosity function down to $M_{\rm UV}=-18.5$, whereas \citet{tanaka_discovery_2025} integrate down to $M_{\rm UV}=-20$. We also show, in red, the results obtained using stricter $\mathrm{F277W\text{--}F444W}$ color cuts, which closely align with the expected evolution proposed by \citet{inayoshi_little_2025}.

Finally, when extrapolated to lower redshifts, our measurements are consistent with results at $z\approx2$ reported by \citet{ma_counting_2025} and \citet{bisigello_euclid_euclid_2025}. However, such extrapolations should be treated with caution, as selection effects can strongly impact these comparisons. Improved constraints on the LRD population toward Cosmic Noon will require future facilities such as the {\it Roman} Space Telescope, which will enable a more complete census beyond current JWST capabilities.

\section{Discussion and Summary}
The advent of JWST has rapidly transformed our view of the high-redshift Universe, uncovering a new population of compact red sources, commonly referred to as LRDs. Significant effort has been devoted to constraining their physical nature (e.g., \citealt{naidu_black_2025, de_graaff_remarkable_2025, pacucci_cosmic_2025, inayoshi_little_2025, juodzbalis_jades_2024, juodzbalis_jades_2025, ji_blackthunder_2025, barro_cliff_2025, perez-gonzalez_little_2026, madau_little_2026}) and, more recently, to exploring their possible evolutionary pathways at later cosmic times (e.g., \citealt{billand_investigating_2025, rinaldi_beyond_2025}). As the number of LRD studies has increased, photometric selection strategies have become increasingly refined and now constitute the primary route for assembling large samples of candidates for spectroscopic follow-up.

Over time, several studies have produced the first LRD catalogs across the main JWST legacy fields. For example, \citet{akins_cosmos-web_2025} presented a census from the COSMOS-Web survey, identifying 434 sources over 0.54 deg$^{2}$, while \citet{kokorev_census_2024} and \citet{kocevski_rise_2025} performed selections across multiple legacy fields, including CEERS, PRIMER, and JADES, each covering approximately 0.17 deg$^{2}$ and yielding samples of 260 and 341 LRDs, respectively. Although COSMOS-Web provides the largest JWST area with NIRCam and MIRI coverage, it remains among the shallowest datasets, whereas surveys such as CEERS and PRIMER reach greater depths over much smaller areas. In contrast, the JADES fields now place us in a qualitatively different regime by combining a large area with the deepest NIRCam imaging available to date on such scales, thanks to the latest DR5. Motivated by this, we present an updated (and more inclusive) census of LRDs in the JADES fields, providing a uniform and up-to-date reference sample for future physical studies and targeted spectroscopic follow-up.

\subsection{Are we missing the majority of Little Red Dots?}

Using the full depth and homogeneity of the JADES DR5 imaging, we make a comprehensive census of LRDs across GOODS-N and GOODS-S, spanning $z\approx2$ to $z\approx11$. Our selection balances inclusiveness and purity by combining relaxed color criteria ($\mathrm{F150W\text{--}F200W}<1$~mag and $\mathrm{F277W\text{--}F444W}>0.5$~mag) with a stringent compactness requirement based on direct F444W size measurements ($R_{\rm eff,F444W}<0.06$\arcsec).

We find that strong emission-line galaxies are the primary source of contamination in photometric LRD selections, particularly at $z \gtrsim 6\text{--}7$, where NIRCam photometry alone cannot robustly distinguish a rising continuum. In this regime, MIRI data are required to isolate genuine LRDs, while visual inspection of their SEDs remains necessary to mitigate spurious identifications based on NIRCam alone. Consequently, the LRD population at high redshift remains poorly constrained, given the limited depth and area of existing MIRI surveys. A clear example is {\it Virgil} ($z\approx6.6$; \citealt{iani_midis_2024, rinaldi_deciphering_2025}), which fails standard NIRCam-based selections but is recovered when MIRI data are included (see also \citealt{barro_cliff_2025}).

These limitations directly affect estimates of the LRD number-density evolution at $z \gtrsim 6\text{--}7$, where both observations and models indicate a rapid decline toward early times (e.g., \citealt{inayoshi_little_2025, tanaka_discovery_2025}). While this trend has been attributed to a combination of intrinsic evolution of LRDs and observational effects (such as cosmological surface-brightness dimming ; see \citealt{billand_investigating_2025, pacucci_cosmic_2025, rinaldi_beyond_2025}), it may also arise from our current observational limitations. In particular, the scarcity of deep and wide MIRI data\footnote{Deep and ultra-deep MIRI observations currently cover very limited areas; e.g., MIDIS (\citealt{ostlin_miri_2024}).} limits the identification of LRDs at $z\gtrsim6\text{--}7$, where NIRCam alone increasingly samples only the rest-frame UV. As a result, sources that would qualify as LRDs with sufficiently deep mid-infrared data may remain undetected or misclassified. The observed decline at high redshift may therefore reflect incompleteness as much as intrinsic evolution, leaving their relative contributions uncertain. The fraction of LRDs currently missed at these redshifts is essentially unconstrained, and consequently the epoch at which LRDs first emerge in the early Universe remains poorly determined.

We find that photometrically selected LRDs are found to be extremely compact, with a median F444W effective radius of $R_{\rm eff}\approx125$~pc, well below the typical sizes of high-redshift star-forming galaxies at high redshifts \citep[e.g.,][]{langeroodi_evolution_2023, ormerod_epochs_2024}, and fully consistent with recent LRD size measurements \citep{kokorev_census_2024, baggen_small_2024, furtak_jwst_2023}. Moreover, \citet{whalen_limitations_2025} have shown that size estimates obtained with {\sc pysersic} tend to be biased high for LRDs, implying that our photometrically selected LRDs are intrinsically even more compact than inferred here. Together, these results indicate that the optical compactness is a fundamental selection criterion, as also suggested by \citet{hviding_rubies_2025}. 

Finally, the $\mathrm{F277W\text{--}F444W}$ color distribution shows that very red sources (e.g., $\gtrsim 1.5$ mag; \citealt{akins_cosmos-web_2025}) constitute only a minor fraction of the LRD population ($\lesssim 25\%$), consistent with recent spectroscopic results (\citealt{perez-gonzalez_little_2026}). The majority instead occupy significantly bluer optical/NIR colors (e.g., $\approx 55$\% with $\rm F277W\text{--}F444W = 0.5\text{--}1$~mag). This region of parameter space remains largely unexplored, likely biasing current samples toward an extreme subset of LRDs \citep{perez-gonzalez_little_2026} rather than the full population. If current selection criteria, and subsequent follow-up efforts, are preferentially sensitive to the reddest systems, then our current view of LRDs is inherently biased toward a specific sub-population, preventing both (i) a meaningful inference of their evolutionary trends across cosmic time and (ii) a robust physical interpretation of the LRD phenomenon, by disproportionately emphasizing a single sub-type of a broader, heterogeneous population (e.g., \citealt{barro_cliff_2025, perez-gonzalez_little_2026}), potentially favoring ad hoc models and hindering the development of a unified framework that captures their full diversity and possible evolutionary pathways.

\subsection{The role of selection criteria in Little Red Dot evolution}

We compute the UV luminosity function of photometrically selected LRDs across four redshift bins spanning $z \approx 2$ to $z \approx 10.5$ (approximately 2.7 Gyr in cosmic time).  At lower redshifts ($z \approx 2\text{--}4.5$), where current JWST constraints remain limited, two trends emerge. First, our more inclusive selection yields number densities systematically higher (particularly at faint $M_{\rm UV}$) than those reported by \citet{ma_counting_2025} and \citet{bisigello_euclid_euclid_2025}, regardless of whether we adopt the primary selection alone or include a dedicated low-$z$ search. Second, imposing much stricter red color cuts (e.g., $\gtrsim 1.5$ mag) significantly suppresses the recovery of LRDs in this redshift range. This highlights the intrinsic difficulty of identifying LRDs toward Cosmic Noon, where the choice of filters and color criteria plays a critical role. Nonetheless, any comparison with \citet{ma_counting_2025} and \citet{bisigello_euclid_euclid_2025} should be interpreted with caution, given the differences in selection techniques, observing facilities, and the adopted binning in cosmic time. In this context, the {\it Roman} Space Telescope will represent a key facilities to study LRDs at later cosmic times.

At $z \approx 4.5\text{--}6.5$ and $z \approx 6.5\text{--}8.5$ (the redshift range most extensively explored by the LRD literature so far), our inferred number densities are consistent with recent LRD studies and JWST-selected AGN samples \citep[e.g.,][]{harikane_jwstnirspec_2023, maiolino_jades_2023, greene_uncover_2024, kocevski_rise_2025, kokorev_census_2024, matthee_little_2024, akins_cosmos-web_2025, juodzbalis_jades_2025}, and broadly follow the expected QSO evolution predicted by \citet{li_reconstruction_2024}. In this regime, our results are broadly insensitive to the adopted $\mathrm{F277W\text{--}F444W}$ color threshold.

At the highest redshifts ($z\approx8.5\text{--}10.5$), our UV LF measurements are consistent with \citet{tanaka_discovery_2025} and extend current constraints to fainter luminosities. The inferred number densities lie above predictions for QSOs \citep{zhang_trinity_2024}, but remain well below the UV LF of star-forming galaxies at similar epochs \citep{finkelstein_coevolution_2022}. We note, however, that spectroscopic confirmation is required to robustly establish the nature of this subsample.

\begin{figure}[ht!]
    \centering
    \includegraphics[width=1.0\linewidth]{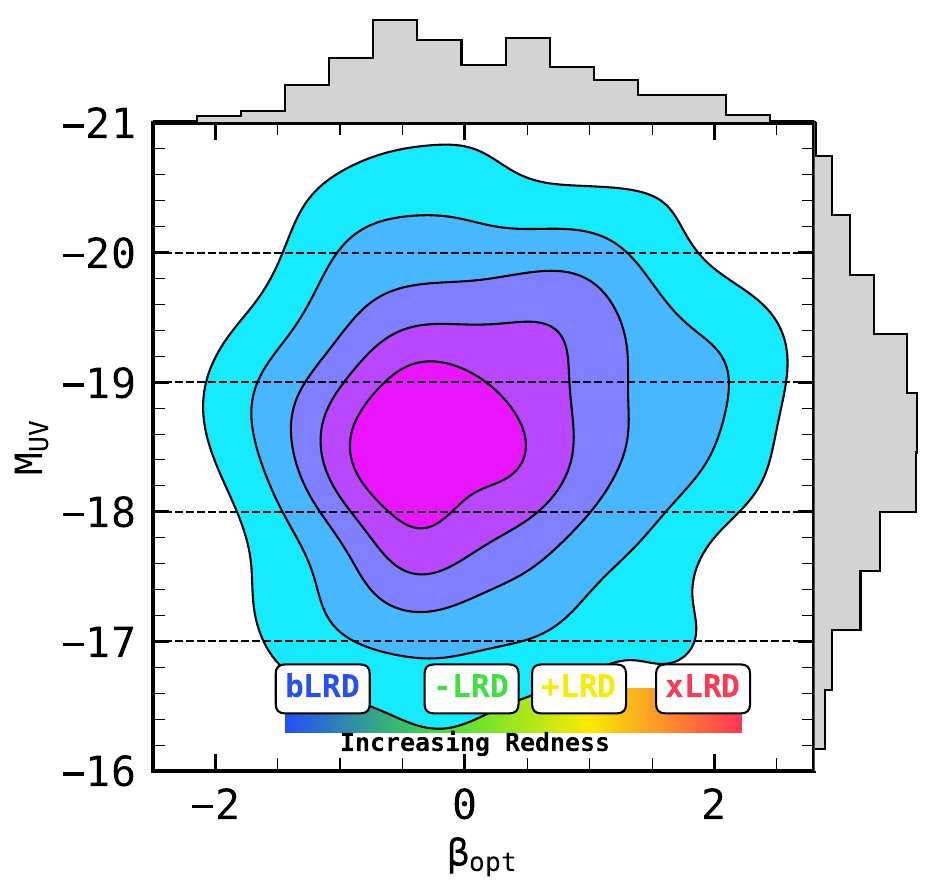}
    \caption{$\beta_{\rm opt}$ versus $M_{\rm UV}$. At fixed $M_{\rm UV}$, sources span a broad range of optical slopes, indicating a wide diversity in redness (also reflected by colors). This demonstrates that a given $M_{\rm UV}$ samples multiple LRD “flavors”, from less extreme to very red systems.}
    \label{fig:MUV_vs_slope_opt}
\end{figure}

However, examining the UV LF of LRDs is non-trivial, as at fixed $M_{\rm UV}$ it likely mixes different LRD “flavors” (\citealt{perez-gonzalez_little_2026}). This is evident in Figures~\ref{fig:L5100_beta_opt} as well as \ref{fig:MUV_vs_slope_opt}, the latter showing that at, fixed $M_{\rm UV}$, LRDs span a broad and continuous range of optical redness (traced by $\beta_{\rm opt}$ or colors), with no significant correlation ($\rho = -0.07$). This implies that similar UV luminosities correspond to diverse spectral configurations (or “flavors”; \citealt{perez-gonzalez_little_2026}), so the UV LF necessarily mixes systems with different physical properties. Nonetheless, even exploring the bolometric LF carries comparable uncertainties as discussed recently in \citet{greene_what_2026}.

For this reason, we instead examine the optical LF as a function of cosmic time. In the lowest redshift bin, our measurements broadly agree with \citet{ma_counting_2025} at the bright end, while diverging toward lower luminosities, where they remain consistent with the quasar optical LF and lie below the observed SFG population (\citealt{ross_sdss-iii_2013, marchesini_evolution_2012}). At higher redshifts, the available literature remains limited; nevertheless, our results agree well with recent measurements by \citet{greene_what_2026} and with de-corrected estimates from \citet{kokorev_census_2024} (interpreted as order-of-magnitude constraints) at intermediate redshifts ($z \approx 4.5\text{--}8.5$), and are broadly consistent with \citet{tanaka_discovery_2025}, the only available constraint at $z \gtrsim 8.5$.

Overall, the optical LF shows little evolution in number density at the bright end for $z\gtrsim4.5$, where our sample is likely most complete. At progressively lower $L_{5100}$, we do observe variations with cosmic time; however, given that our analysis is limited to only the JADES fields, these trends remain poorly constrained. In contrast, the lowest redshift bin shows a more pronounced decline. Whether this reflects intrinsic evolution of the LRD population or limitations in identifying these systems with JWST alone remains unclear. In this context, future facilities such as the {\it Roman} space telescope will be critical to robustly trace the evolution of LRDs at $z<4.5$.

We then examine the redshift evolution of the integrated number density of LRDs and compare our measurements with recent observational and theoretical expectations. To do so, we adopt the same framework recently explored by \citet{inayoshi_little_2025} and \citet{pacucci_cosmic_2025}. We find that the inferred evolution depends sensitively on the adopted degree of redness. A more inclusive selection yields higher number densities at $z \lesssim 6$ and a relatively smooth evolution with cosmic time when compared with current predictions from \citet{inayoshi_little_2025} and \citet{pacucci_cosmic_2025}. In contrast, restricting the sample to extremely red sources ($\rm F277W\text{--}F444W > 1.5$~mag) results in lower number densities and a sharper decline toward lower redshifts, in closer agreement with theoretical predictions by \citet{inayoshi_little_2025} at $z \lesssim 4$.

This behavior reflects the fact that different selection strategies can probe different regions of the LRD parameter space. Indeed, the extremely red subsample represents only a small fraction of the overall population (as already noted in \citealt{perez-gonzalez_little_2026}), and isolating it effectively selects a subset with a distinct redshift distribution. Conversely, relaxing the color threshold includes sources with less extreme optical slopes, yielding a higher normalization and a broader evolution. As a result, the inferred number-density evolution of LRDs is not unique, but depends directly on how restrictive the adopted color selection is; i.e., on the degree of redness used to define the population observationally. 

Importantly, we observe that $\beta_{\rm opt}$ (and similarly $\rm F277W\text{--}F444W$ or $\rm F200W\text{--}F444W$) increases with redshift, while $L_{5100}$ shows a weaker positive trend. This indicates that higher-$z$ LRDs are progressively characterized by redder spectral slopes, with luminosity playing a secondary role. When integrating the UV LF down to a fixed limit to estimate $n(z)$ as a function of cosmic time (e.g., $M_{\rm UV}=-18.5$, as in our case), this implies that different subsets of the LRD population are effectively sampled at different redshifts: at low $z$ the counts can include a broad range of LRD sub-types, while at high $z$ they are increasingly dominated by the reddest systems (which tend to be the brightest). Consequently, the measured $n(z)$ does not trace a single homogeneous population, but likely rather a redshift-dependent mixture of objects occupying different regions of the color–luminosity space. We stress, however, that selection effects may play an important role here; indeed, at $z \gtrsim 6\text{--}7$, the lack of deep MIRI data limits our ability to identify LRDs (see also discussion in \citealt{barro_cliff_2025}), biasing the sample toward the reddest sources, while at lower redshifts the specific choice of filters used to define color criteria can also affect which objects are classified as LRDs.

Therefore, within this framework, different photometric selections sample distinct subsets of the intrinsically heterogeneous LRD population. In particular, restricting the selection to the reddest systems isolates sources with the most extreme optical slopes, which can significantly alter the inferred evolution of LRDs by skewing their redshift distribution toward higher $z$ (Figure~\ref{fig:LRD_evo_z}). Adopting more inclusive criteria reveals a more extended LRD population spanning a continuous range in optical redness, from less to extremely red systems (in agreement with recent stacking analyses of spectroscopically confirmed LRDs; \citealt{perez-gonzalez_little_2026}), and yields a smoother inferred evolution compared to studies based on stringent red cuts (e.g., \citealt{inayoshi_little_2025}). 

If LRDs represent a heterogeneous population spanning a continuum from less red to extremely red systems, potentially associated with different physical properties (e.g., apparent Balmer-like breaks, emission line properties, and varying host galaxy fractions; see, e.g., \citealt{de_graaff_little_2025, ji_blackthunder_2025, perez-gonzalez_little_2026, barro_cliff_2025, matthee_engine_2026, sun_little_2026}), then their evolution with cosmic time must be interpreted with caution. It remains unclear whether distinct evolutionary trends can be robustly identified for different sub-populations, given the apparent continuity from less red to extremely red LRDs. In this context, current constraints on the global evolution of LRDs are likely biased toward a specific, yet still mixed, sub-branch of this broader population; moreover, analyses that integrate down to a fixed $M_{\rm UV}$ at a given epoch inherently sample different portions of the population, further complicating the interpretation of their evolution with cosmic time.

\subsection{Final remarks}
To date, much of the observational effort on LRDs has focused on their most extreme manifestations, implicitly treating the reddest objects as representative of the population as a whole. However, when adopting a more inclusive selection, we find instead that LRDs span a wide and continuous range of colors, optical luminosities, SED shapes, and inferred physical properties, reinforcing recent results indicating that this class is intrinsically heterogeneous rather than a narrowly defined phenomenon (e.g., \citealt{barro_cliff_2025, perez-gonzalez_little_2026}). 

Our results highlight that current LRD samples are strongly biased toward a specific sub-population, and that both low- and high-redshift searches remain vulnerable to systematic selection effects, which in turn can heavily bias our understanding of their evolution as a function of cosmic time. In particular, at $z \gtrsim 6\text{--}7$, NIRCam photometry alone cannot reliably trace the rising continuum redward of the Balmer break, making the identification of genuine LRDs increasingly uncertain and perhaps biasing our census of LRDs in the early Universe. Fully characterizing the LRD phenomenon requires moving beyond extremely red sources and exploring the full degree of redness exhibited by LRDs. The large and homogeneous sample presented in this work provides a critical foundation for future spectroscopic surveys. 

Finally, we stress that deep, wide-area MIRI observations are fundamental to trace the LRD population into the epoch where their first emerge ($z\gtrsim8$), as NIRCam alone is insufficient at these redshifts. The LRD candidate at $z\approx10$ reported by \citet{tanaka_discovery_2025} confirms their presence in the early Universe, but probes only the bright end ($L_{5100} \gtrsim 10^{44}\,\mathrm{erg\,s^{-1}}$), corresponding to the extreme, and intrinsically rare (\citealt{perez-gonzalez_little_2026}), branch of the LRD population. As a result, current shallow MIRI surveys are insufficient to study LRDs, as they access only a small and biased fraction of the overall population in the early Universe.
\vspace{4mm}

\acknowledgments

We thank Edoardo Iani, Rohan Naidu, Fabio Pacucci, Vasily Kokorev, and Jorryt Matthee for useful discussions.
The final catalog will be made publicly available upon acceptance of this paper.

This work is based on observations made with the NASA/ESA/CSA JWST. The data were obtained from the Mikulski Archive for Space Telescopes at the Space Telescope Science Institute, which is operated by the Association of Universities for Research in Astronomy, Inc., under NASA contract NAS5-03127 for JWST. These observations are associated with PIDs 1063, 1176, 1180, 1181, 1210, 1264, 1283, 1286, 1287, 1895, 1963, 2079, 2198, 2514, 2516, 2674, 3215, 3577, 3990, 4540, 4762, 5398, 5997, 6434, and 6511. The authors acknowledge the teams of programs 1895, 1963, 2079, 2514, 3215, 3577, 3990, 6434, and 6541 for developing their observing program with a zero-exclusive-access period. The specific observations analyzed can be accessed via doi:10.17909/8tdj-8n28. Additionally, this work made use of the lux supercomputer at UC Santa Cruz, which is funded by NSF MRI grant AST1828315, as well as the High Performance Computing (HPC) resources at the University of Arizona, which is funded by the Office of Research Discovery and Innovation (ORDI), Chief Information Officer (CIO), and University Information Technology Services (UITS).

PR acknowledges support by JWST/NIRCam contract to the University of Arizona, NAS5-02105.

SA acknowledges support from the JWST Mid-Infrared Instrument (MIRI) Science Team Lead, grant 80NSSC18K0555, from NASA Goddard Space Flight Center to the University of Arizona.

WMB gratefully acknowledges support from DARK via the DARK fellowship. This work was supported by a research grant (VIL54489) from VILLUM FONDEN.

AJB acknowledges funding from the "FirstGalaxies" Advanced Grant from the European Research Council (ERC) under the European Union’s Horizon 2020 research and innovation programme (Grant agreement No. 789056).

SC and GV acknowledge support by European Union’s HE ERC Starting Grant No. 101040227 - WINGS.

CC acknowledges support from the JWST/NIRCam Science Team contract to the University of Arizona, NAS5-02105, and JWST Programs 3215 and 5015.

FDE acknowledges support by the Science and Technology Facilities Council (STFC), by the ERC through Advanced Grant 695671 ``QUENCH'', and by the UKRI Frontier Research grant RISEandFALL.

JWST/NIRCam contract to the University of Arizona NAS5-02105.

DJE is supported as a Simons Investigator and by JWST/NIRCam contract to the University of Arizona, NAS5-02105.  Support for program \#3215 was provided by NASA through a grant from the Space Telescope Science Institute, which is operated by the Association of Universities for Research in Astronomy, Inc., under NASA contract NAS 5-03127.

JMH. is supported by the JWST/NIRCam Science Team contract to the University of Arizona, NAS5-02105, along with JWST Programs 3215 and 8544.

XJ and IJ acknowledges support by the Science and Technology Facilities Council (STFC), by the ERC through Advanced Grant 695671 ``QUENCH'', and by the UKRI Frontier Research grant RISEandFALL.

ZJ acknowledges support by JWST/NIRCam contract to the University of Arizona, NAS5-02105.

RM acknowledges support by the Science and Technology Facilities Council (STFC), by the ERC through Advanced Grant 695671 “QUENCH”, and by the UKRI Frontier Research grant RISEandFALL. RM also acknowledges funding from a research professorship from the Royal Society.

PGP-G acknowledges support from grant PID2022-139567NB-I00 funded by Spanish Ministerio de Ciencia e Innovaci\'on MCIN/AEI/10.13039/501100011033, FEDER, UE.

ST acknowledges support by the Royal Society Research Grant G125142.

BER acknowledges support from the NIRCam Science Team contract to the University of Arizona, NAS5-02105, and JWST Program 3215.

JS acknowledges support by the Science and Technology Facilities Council (STFC), ERC Advanced Grant 695671 "QUENCH".

RM acknowledges support by the Science and Technology Facilities Council (STFC), by the ERC through Advanced Grant 695671 “QUENCH”, and by the UKRI Frontier Research grant RISEandFALL. RM also acknowledges funding from a research professorship from the Royal Society.

BER acknowledges support from the NIRCam Science Team contract to the University of Arizona, NAS5-02015, and JWST Program 3215.

The research of CCW is supported by NOIRLab, which is managed by the Association of Universities for Research in Astronomy (AURA) under a cooperative agreement with the National Science Foundation.

JW gratefully acknowledges support from the Cosmic Dawn Center through the DAWN Fellowship. The Cosmic Dawn Center (DAWN) is funded by the Danish National Research Foundation under grant No. 140.

IJ acknowledges support by the Huo Family Foundation through a P.C. Ho PhD Studentship.

\appendix
\section{Examples of photometrically selected LRDs}\label{appendix_example}

\begin{figure*}[ht!]
    \centering
    \includegraphics[width=1.0\linewidth]{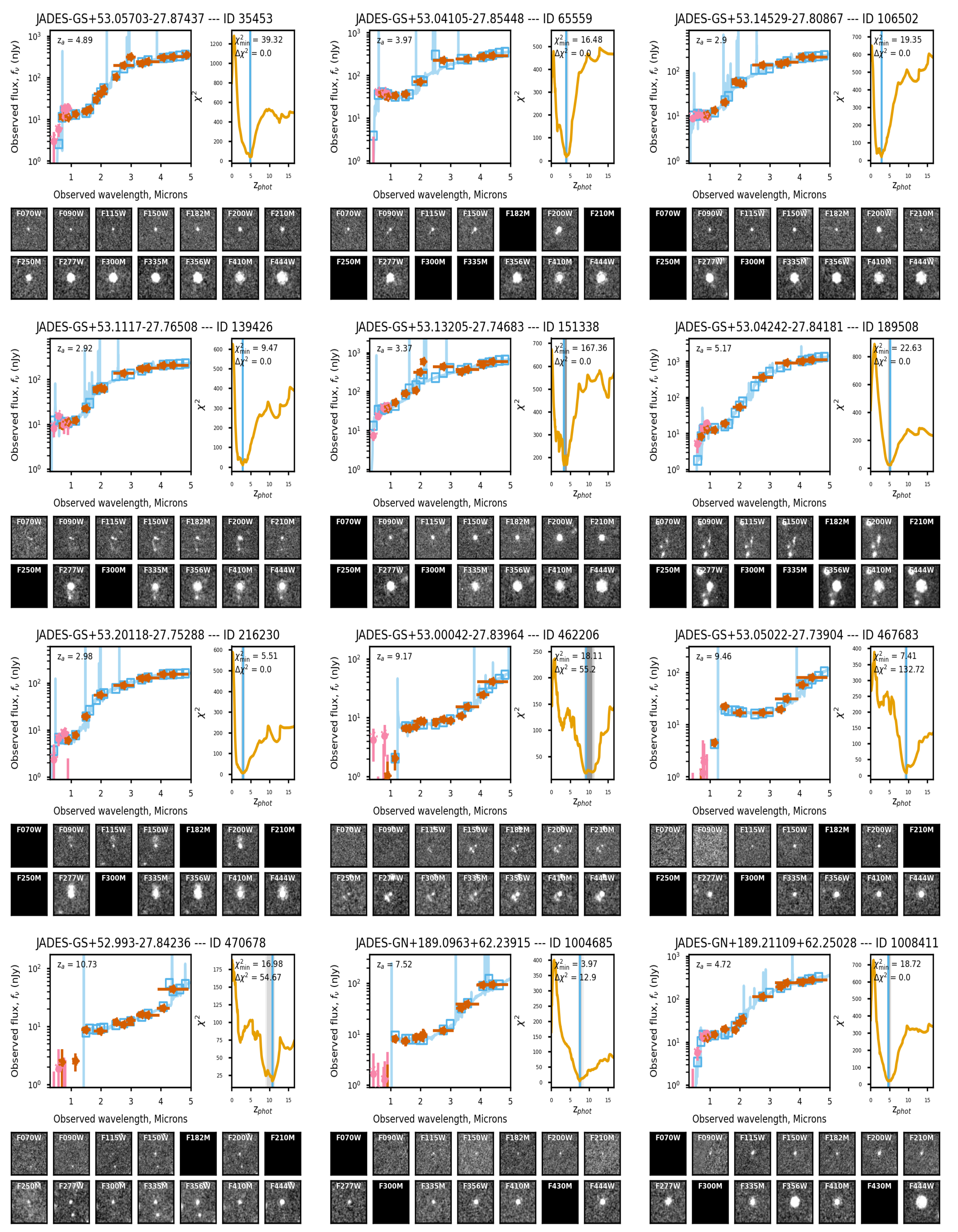}
    \caption{Examples of SEDs and postage stamps for LRDs across the full redshift range of our sample, from low to high $z$. 
}  
    \label{fig:example_LRDs}
\end{figure*}

\bibliography{references}{}
\bibliographystyle{aasjournal}

\end{document}